\documentclass[twocolumn,prc,aps,showkeys,superscriptaddress,nofootinbib]{revtex4}

\usepackage{amssymb}
\usepackage{amsmath}
\usepackage{graphicx} % Include figure files
\usepackage{epsfig}   % Include figure files
\usepackage{dcolumn}  % Align table columns on decimal point
\usepackage{rotating} 
\usepackage[dvips]{color}

\begin{document}

%\centerline{\it \ham {\huge\bf DRAFT-V8} }
%\vspace{0.2in}

%%%%%%%%%%%%%%%%%%%%%

\title{The lead-glass electromagnetic calorimeters for the magnetic spectrometers in Hall C at 
Jefferson Lab}

\newcommand*{\ANSL}{A.~I.~Alikhanyan National Science Laboratory (Yerevan Physics Institute), 
Yerevan 0036, Armenia}

\newcommand*{\JLAB}{Thomas Jefferson National 
Accelerator Facility, Newport News, Virginia 23606, USA}

\newcommand*{\HAMPTON}{Hampton University, Hampton, Virginia 23668, USA }

\newcommand*{\ARGONNE}{Physics Division, Argonne National Laboratory, Argonne, Illinois 60439, USA }

\newcommand*{\CALTECH}{California Institute of Technology, Pasadena, California 91125, USA }

\newcommand*{\DUKE}{Triangle Universities Nuclear Laboratory and Duke University, 
Durham, North Carolina 27708, USA }

\newcommand*{\FIU}{Florida International University, University Park, Florida 33199, USA }

\newcommand*{\GETTY}{Gettysburg College, Gettysburg, Pennsylvania 18103, USA}

\newcommand*{\GWU}{The George Washington University, Washington, D.C. 20052, USA }

\newcommand*{\HOU}{University of Houston, Houston, TX 77204, USA }

\newcommand*{\JMU}{James Madison University, Harrisonburg, Virginia 22807, USA }

\newcommand*{\MARYLAND}{University of Maryland, College Park, Maryland 20742, USA }

\newcommand*{\MSS}{Mississippi State University, Mississippi State, Mississippi 39762, USA }

\newcommand*{\NCSU}{North Carolina A \& T State University, Greensboro, North Carolina 27411, USA}

\newcommand*{\OHIO}{Ohio University, Athens, Ohio 45071, USA }

\newcommand*{\REGINA}{University of Regina, Regina, Saskatchewan, S4S 0A2, Canada }

\newcommand*{\RPI}{Rensselaer Polytechnic Institute, Troy, New York 12180, USA}

\newcommand*{\RUTGERS}{Rutgers, The State University of New Jersey, 
Piscataway, New Jersey, 08855, USA }

\newcommand*{\UCONN}{University of Connecticut, Storrs, Connecticut 06269, USA }

\newcommand*{\UMASS}{University of Massachusetts Amherst, Amherst, Massachusetts 01003, USA}

\newcommand*{\UNC}{University of North Carolina Wilmington, Wilmington, North Carolina 28403, USA }

\newcommand*{\UVA}{University of Virginia, Charlottesville, Virginia 22901, USA }

\newcommand*{\VASS}{Vassar College, Poughkeepsie, New York 12604, USA }

\newcommand*{\VU}{VU-University, 1081 HV Amsterdam, The Netherlands }

\newcommand*{\JOHAN}{University of Johannesburg, Johannesburg, South Africa }

\newcommand*{\WITS}{University of the Witwatersrand, Johannesburg, South Africa }

\newcommand*{\CUA}{Catholic University of America, Washington, DC 20064}

\newcommand*{\LNSMIT}{Laboratory for Nuclear Science. Massachusetts Institute of Technology, 
Cambridge, MA, USA}

\author{H.~Mkrtchyan}
\affiliation{\ANSL}

\author{R.~Carlini}
\affiliation{\JLAB}

\author{V.~Tadevosyan}
%%\thanks{Corresponding author. Tel: + 357 269 7741.}
%%\email{tadevosn@jlab.org}
\affiliation{\ANSL}

\author{J.~Arrington} 
\affiliation{\ARGONNE}

\author{A.~Asaturyan}
\affiliation{\ANSL}

\author{M.~E.~Christy} 
\affiliation{\HAMPTON}

\author{D.~Dutta}
\affiliation{\DUKE}

\author{ R.~Ent}
\affiliation{\JLAB}

\author{H.~C.~Fenker}
\affiliation{\JLAB}

\author{D.~Gaskell}
\affiliation{\JLAB}

\author{T.~Horn} 
\affiliation{\CUA}

\author{M.~K.~Jones}
\affiliation{\JLAB}

\author{C.~E.~Keppel} 
\affiliation{\HAMPTON}

\author{D.~J.~Mack}
\affiliation{\JLAB}

\author{S.~P.~Malace}
\affiliation{\DUKE}

\author{A.~Mkrtchyan}
\affiliation{\ANSL}

\author{M.~I.~Niculescu}
\affiliation{\JMU}

\author{J.~Seely}
\affiliation{\LNSMIT}

\author{V.~Tvaskis}
\affiliation{\HAMPTON}

\author{S.~A.~Wood}
\affiliation{\JLAB}

\author{S.~Zhamkochyan}
\affiliation{\ANSL}

\newpage
\date{\today}

%%%%%%%%%%%%%%%%%%%%%%%%%%%%%%%%%%%%%%%%%%%%%%%%%%%%%%%%%%%%%%%%%%%%%%%%%%%%%%%%%%%%%%%%%%%%%%%%%%%

\begin{abstract}
The electromagnetic calorimeters of the various magnetic spectrometers in Hall C at
Jefferson Lab are presented. For the existing HMS and SOS spectrometers design considerations,
relevant construction information, and comparisons of simulated and experimental results are
included. The energy resolution of the HMS and SOS calorimeters is better than
$\sigma/E \sim 6\%/\sqrt E $, and pion/electron ($\pi/e$) separation of about 100:1
has been achieved in energy range 1 -- 5 GeV. Good agreement has been observed
between the experimental and simulated energy resolutions, but simulations systematically exceed
experimentally determined $\pi^-$ suppression factors by close to a factor of two.
For the SHMS spectrometer presently under construction details on the design and
accompanying GEANT4 simulation efforts are given. The anticipated performance of the new
calorimeter is predicted over the full momentum range of the SHMS. Good electron/hadron separation
is anticipated by combining  the energy deposited in an initial (preshower) calorimeter layer
with the total energy deposited in the calorimeter. 
\end{abstract}



\maketitle

%%%%%%%%%%%%%%%%%%%%%%%%%%%%%%%%%%%%%%%%%%%%%%%%%%%%%%%%%%%%%%%%%%%%%%%%%%%%%%%%%%%%%%%%%%%%%%%%%%%%
\section{Introduction}
\label{intro}

The experimental program at Jefferson Lab focuses on the studies of the electromagnetic structure of 
nucleons and nuclei, in particular in a region where a transition is expected from a nucleon-meson
description into a quark-gluon description of matter. In experimental Hall C the emphasis has been
on inclusive (e,e$^\prime$) electron scattering and proton knockout (e,e$^\prime$p) experiments
at the highest four-momentum transfer ($Q^2$) accessible, deuteron photodisintegration experiments,
and both exclusive and semi-inclusive pion electroproduction reactions. In particular, the Hall C
experimental program has studied the onset of the quark-parton model description of such reactions.
To accomplish such a diverse program, a highly flexible set of instruments capable of accurate 
measurements of final momenta and angles is required, including both efficient background rejection
and good particle identification properties. This remains very much in place after the 12-GeV
Upgrade of Jefferson Lab (JLab) has been completed, with Hall C emphasizing precision measurements at
high luminosities, with detection of high-energy reaction products approaching the beam energy at
very forward angles.

The initial base equipment of Hall C was well suited to the JLab scientific program that required
high luminosity, intermediate detector acceptances and resolution \cite{CDR1990}.
With the high luminosities needed to access neutrino-like scattering
probabilities comes a high-background suppression requirement. The magnetic spectrometer pair that
constituted the base equipment pointed to a common pivot with scattering chamber. The Short Orbit 
Spectrometer (SOS), with a QD\=D configuration, accessed a momentum range of 0.3 - 1.7 GeV/c, and
an angular range of $13.3^{\circ}$ - $168.4^{\circ}$. It was explicitly designed to measure
pions and kaons with short life times. The High Momentum Spectrometer (HMS), with a QQQD magnetic 
configuration, covered a momentum range 0.5 - 7.3 GeV/c, but was to date not used above 5.7
GeV/c. The HMS accessed an angular range between $10.5^{\circ}$ - $80^{\circ}$.

After the JLab 12-GeV Upgrade \cite{CDR-12}, the Hall C scientific program is again focused on
high luminosity measurements with detection of high energy reaction products at small forward angles.
Such a physics program can be accessed only by a spectrometer system providing high acceptance for,
given the larger boosts associated with the energy upgrade, very forward-going particles, and
analyzing power for particle momenta approaching that of the incoming beam.
To accomplish this, and maintain a spectrometer pair rotating around a common pivot for
precision coincidence measurements, the SOS will be superseded by the newly built
Super High Momentum Spectrometer (SHMS). The SHMS will achieve a minimum (maximum) scattering angle
of 5.5$^\circ$ (40$^\circ$) with acceptable solid angle and do so at high luminosity. The maximum
momentum will be 11 GeV/c, well matched to the maximum energy available in Hall C. 
The basic parameters of the HMS, SOS and SHMS are listed in Table~\ref{hms-shms-param}.

\begin{center}
\begin{table}
\caption{\label{hms-shms-param} The basic parameters of the HMS, SOS and SHMS spectrometers.}
{\centering  \begin{tabular}{|c|c|c|c|}
\hline
Parameter                       &  HMS          & SOS            &   SHMS       \\
\hline
Momentum Range (GeV/c)          & 0.5-7.3       & 0.3-1.7        & 1.5-11.0      \\ 
Momentum Acceptance (\%)        & $\pm 10$      & $\pm 20$       & -10 - +22    \\
Momentum resolution (\%)        & 0.10-0.15     & $<$0.1         & 0.03-0.08    \\
Horiz. Angl. Accept.(mrad)      & $\pm$32       & $\pm$40        & $\pm$18      \\
Vert. Angl. Accept. (mrad)      & $\pm$85       & $\pm$70        & $\pm$50      \\
Solid angle (msr)               & 8.1           & 9.0            & $>$ 4.5      \\
Maximum scattering angle        & $\leq 80^o$   & $\leq 168.4^o$ & $\leq 40^o$  \\
Minimum scattering angle        & $\geq 10.5^o$ & $\geq 13.3^o$  & $\geq 5.5^o$ \\
Horiz. Angl. res. (mrad)        & 0.8           & 0.5            & 0.5-1.2      \\
Vertical  Angl. res. (mrad)     & 1.0           & 1.0            & 0.3-1.1      \\
Vertex Reconstr. res. (cm)      & 0.3           & 2-3            & 0.1-0.3      \\
\hline
\end{tabular}\par}
\end{table}
\end{center}

\begin{figure}
\begin{center}
\epsfxsize=3.40in
\epsfysize=1.70in
\epsffile{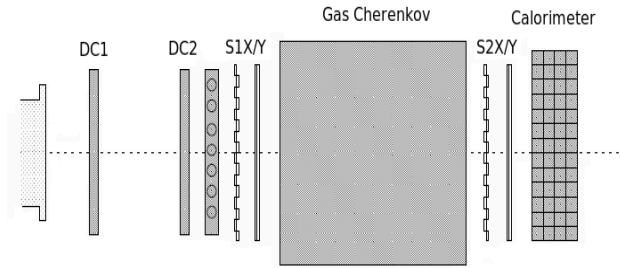}
\caption{\label{hms_hut}
Schematic side-view of the HMS detector package. The aerogel detector was added between the DC2
drift chamber and the S1X hodoscope in 2003.
Adapted from \cite{arrington}.}
\end{center}
\end{figure}

The standard detector packages in the HMS and SOS were designed from inception to be very similar
\cite{arrington}.
The detector stacks, shown for the HMS in Fig.~\ref{hms_hut}, are located inside the respective 
concrete spectrometer shield houses. A pair of six-plane drift chambers (DC1 and DC2) is 
situated immediately after the dipole magnet, in the forefront of shield house to allow for 
particle tracking. They are followed by two pairs of x-y scintillator hodoscopes 
sandwiching a gas \v{C}erenkov. In some experiments, an aerogel \v{C}erenkov detector was added 
either before (HMS) or after (SOS) the pairs of scintillators.
The last detector in the detection stack is the lead-glass electromagnetic calorimeter, positioned
at the very back of the shield house. Its support structure is in fact mounted on the concrete
wall of the shield house. The two sets of drift chambers are used for track reconstruction, the four 
scintillating hodoscope arrays for triggering and time-of-flight measurements, and the threshold 
gas (and aerogel) \v{C}erenkov detectors and lead-glass calorimeters for electron/hadron separation.

Hall C experiments typically demand well-understood detection efficiencies of better than 99\%, and
background particle suppression of 1,000:1 in $e/\pi$ separation, typically. This can be achieved by 
combining 100:1 suppression in the electromagnetic calorimeter, with the remaining 
suppression in a gas \v{C}erenkov counter. Several experiments used signals from the calorimeter and 
gas \v{C}erenkov counters already in a hardware trigger to reject pions or electrons by a factor of 
25:1 in the online data acquisition system. 
The Particle Identification (PID) systems of both spectrometers performed remarkably stable over 
more than a decade of use. The combination of gas \v Cerenkov and lead-glass 
electromagnetic calorimeter ensured pion suppressions of typically a few 1,000:1, for electron 
detection efficiency of better than 98\%.

The detector package of the SHMS will be a near-clone of the HMS. It will again include a pair of 
multiwire drift chambers for tracking, and scintillator and quartz hodoscopes for timing.
As the SHMS will both detect a variety of hadrons ($\pi$,$K$,p) in a number of coincidence
experiments with HMS, and electrons in single-arm (e,e$^\prime$) experiments, special attention is
again paid to the PID system.
It must provide similar particle identification as mentioned above, even at the higher 
energies. In its basic configuration the SHMS detection stack includes a heavy gas \v Cerenkov for 
hadron selection, and a noble-gas \v Cerenkov and lead-glass electromagnetic calorimeter for 
electron/hadron separation. 
It is again envisaged to augment the detector stack with aerogel \v Cerenkov detectors, primarily for
kaon identification. The approved experiments demand a suppression of pion background for 
electron/hadron separation of 1,000:1, with suppression in the electromagnetic calorimeter alone on 
the level of 100:1.
An experiment to measure the pion form factor at the highest $Q^2$ accessible at JLab with 11 GeV 
beam~\cite{Fpi-12} requires a strong suppression of electrons against negative pions of a few 
1,000:1, with a requirement on the electromagnetic calorimeter of a 200:1 suppression.

This paper describes the electromagnetic calorimeters in the various magnetic spectrometers, be it
existing or under construction, in Hall C at Jefferson Lab. Section~\ref{hcal_scal} describes in 
detail the pre-assembly studies, the component selection, construction and assembly of the HMS and 
SOS calorimeters.
Section~\ref{early_mc} explains the Monte Carlo simulation package used, and highlights the 
structure and some details of the simulation software. Sections~\ref{electronic_calibr} and
~\ref{hcal_scal_perform} cover the electronics and calibration of the calorimeters. We present
resolution, efficiency and hadron rejection capability  of the calorimeters in both HMS and SOS,
and compare experimental data with simulation results. In Section~\ref{shms_calo} we describe details
of the newly designed calorimeter for the SHMS, including information on the component selection and
construction. We also present results of pre-assembly component checkout, and the anticipated 
performance of the SHMS calorimeter from simulation studies.

%%%%%%%%%%%%%%%%%%%%%%%%%%%%%%%%%%%%%%%%%%%%%%%%%%%%%%%%%%%%%%%%%%%%%%%%%%%%%%%%%%%%%%%%%%%%%%%%%%%%

\section{HMS and SOS Calorimeters}
\label{hcal_scal}

Particle detection using electromagnetic calorimeters is based on the production of electromagnetic 
showers in a material.
The total amount of the light radiated in this case is proportional to the energy of the primary 
particle.
Electrons (as well as positrons and photons), will deposit their entire energy in the calorimeter 
giving a detected energy fraction of one. The energy fraction is the ratio of energy detected in 
the calorimeter to particle energy.

Charged hadrons entering a calorimeter have a low probability to interact and produce a shower, and 
may pass through without interaction.  In this case they will deposit a constant amount of energy 
in the calorimeter.
However, they may undergo nuclear interactions in the lead-glass and produce particle showers 
similar to the electron and positron induced particle showers. 
Hadrons that interact inelastically near the front surface of the calorimeter and transfer a 
sufficiently large fraction of their energy to neutral pions will mimic electrons. 
The maximum attainable electron/hadron rejection factor is limited mainly by the cross section of 
such interactions.

%---------------------------------------------------------------------------------------------------

\subsection{Construction}
\label{hcal_construct}
R\&D, design and construction of the calorimeters for the HMS and SOS magnetic spectrometers 
started in 1991-1992. 
In 1994 both calorimeters were assembled and installed as part of the instrumentation of 
Hall C spectrometers, becoming the first operational detectors at JLab. 
Since the first commissioning experiment, the calorimeters have been successfully used in nearly
all experiments carried out in Hall C. In 2008, the SOS spectrometer was retired and its 
calorimeter blocks removed to be used for the preshower of the newly designed SHMS spectrometer.   
The HMS calorimeter will remain in place for use after the Continuous Electron Beam Accelerator
Facility's (CEBAF) 12--GeV upgrade. 

\begin{figure}
\begin{center}
\epsfxsize=3.40in
\epsfysize=3.40in
\epsffile{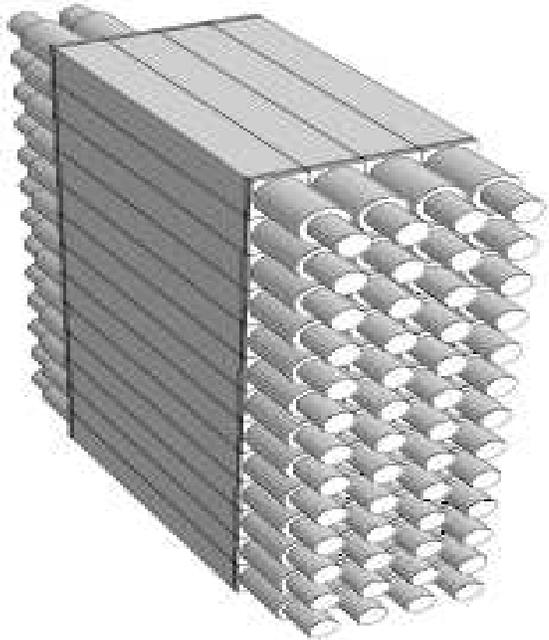}
\caption{\label{hms_lg}
A sketch of the HMS calorimeter. The front of detector is at left.
The left side PMTs were added in 1998.
}
\end{center}
\end{figure}

The HMS/SOS calorimeters are of identical design and construction except for their total size.
Blocks in each calorimeter are arranged in four planes and stacked 13 and 11 blocks high in the HMS
(see Fig.~\ref{hms_lg}) and SOS respectively. 
The planes are shifted relative to each other in the vertical direction by $\sim$5 mm.
In addition, the entire detector is tilted by $5^o$ relative to the central ray of the 
spectrometer. 
These shifts make it  impossible for particles to pass through the calorimeter without interaction.
The total thickness of the material
along the particle direction of $\sim$14.6 radiation lengths is enough to absorb the major part of
energy of electrons within the HMS momentum range.

All blocks were produced in early 1990's by a Russian factory in Lytkarino~\cite{lytkarino}, whose 
products of good optical quality were well known. 
The blocks are $10~{\rm cm}\times 10~{\rm cm}\times 70~{\rm cm}$ in size and machined with a 
precision of 0.05 mm. 
They may contain bubbles or stones with a diameter less than 300 $\mu$m with an impurity frequency
of less than 5-10 per kg of glass.

The optics and acceptances of the spectrometers (see Table~\ref{hms-shms-param})  required 
the calorimeters to have frontal dimensions about $60\times 120~{\rm cm}^2$ for the HMS and 
$60\times 100~{\rm cm}^2$ for the SOS. To avoid any shower leakage from the calorimeter volume, we 
chose to extend the  physical dimensions of the  calorimeters at least 5 cm beyond the sizes 
required by spectrometer acceptance.  This gave calorimeter physical areas of
$70\times 130~{\rm cm}^2$ for the HMS and $70\times 110~{\rm cm}^2$ for the SOS.

Looking from the side, the HMS calorimeter consists of 52 modules stacked in 4 columns (each layer 
3.65 rad. length thick) (see Fig.~\ref{hms_lg}). 
In addition to total energy deposition of the particle, a modular calorimeter gives information 
on the longitudinal development of the shower (which is different for electromagnetic and hadronic 
showers). This additional information can be used for more effective electron/hadron selection.  
Since the modules are oriented transversely to the incident particles, to detect photons from 
\v{C}erenkov radiation one needs to attach photomultipliers (PMTs) from the side of the block
and cover the area $10\times 10~{\rm cm}^2$ of the blocks as effectively as possible.
The energy resolution of a lead-glass shower counter depends
strongly on the ratio of the photocathode area to the output area of
the radiator \cite{davydov}. Photomultipliers with a photocathode diameter of
3.0"-3.5" were considered to be the optimal choice for the HMS/SOS
calorimeters since they could provide a relatively high value of
$\sim$0.44 - 0.50 for this ratio.

%---------------------------------------------------------------------------------------------------

\subsection{The single module assembly}
\label{hcal_module}
\begin{figure}
\begin{center}
\epsfxsize=3.40in
\epsfysize=1.70in
\epsffile{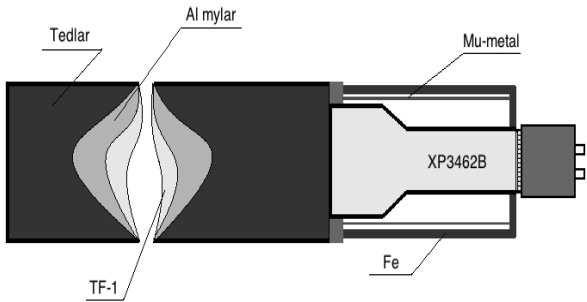}
\caption{\label{calo_block}
Structure of the lead-glass module.}
\end{center}
\end{figure}

The requirement that the lead glass blocks must be optically isolated and optically coupled to PMTs 
was the primary guidance for the construction. 
The individual module design is shown in Fig.~\ref{calo_block}.
To ensure light-tightness, each block is wrapped in 25 $\mu$m thick aluminized Mylar and  
40 $\mu$m thick Tedlar type film. 
There is a thin layer of air between the block and Mylar, for optical insulation was not 
completely tight wrapped.
Each block is also equipped with ST type optical fiber adapter for light monitoring system. 
The blocks are slightly different in sizes, but on average the spread in length is less than 
$\pm$0.250 mm and less than $\pm$0.100 mm (100$\pm$0.10 mm) in transverse size. 
The gaps between the modules in final assembly are less than 250 $\mu$m.

The calorimeter signals from the blocks are read out by 8-stage Philips XP3462B photomultiplier tubes.  
The PMTs are shielded by six turns of 100 $\mu$m thick $\mu$-metal foil.
Since the PMTs operate at negative high voltage and the photocathodes are near the magnetic 
shields and other mechanical parts at ground potential, special protection is required to 
avoid current leakage between the photocathodes and ground. For this reason, the PMT bulbs
were wrapped in several layers of thin Teflon and black electrical tape. 
After full assembly, the current leak for each block was measured with a high voltage about 200 V 
above nominal operating setting.

Silicone grease ND-703 with high viscosity is used for the PMT -- block optical contact
(index of refraction $\sim$1.46).
Originally PMTs were attached to only right side of the blocks (looking along the central ray
of the HMS). 
The PMTs on the left side in the first two layers were added in the late 1998 in order to enhance 
signal output, especially at low energies.
In addition, this weakens dependence of the aggregate signal from a module
on the particle's point of impact.

%---------------------------------------------------------------------------------------------------

\subsection{Photomultiplier tube selection and studies}
\label{hcal-pmt}

The choice of the photomultiplier tube depends on the intensity of light to be measured
and the regime of its operation.
One of the most important requirements for the PMTs used in the HMS/SOS calorimeters was high 
efficiency for the electrons above $\sim$100 MeV, and good linearity up to the energies of several
GeV. 
At low energy (or low light intensity) the PMT must have relatively high gain in order to keep 
electron trigger efficiency high. But, in all cases its operation regime must be 
optimized for best signal-to-noise ratio. 

Ideally, the gain of a PMT with $n$ dynode stages and an average secondary emission ratio 
$\delta$ per stage is $G\sim{\delta}^n$. While the secondary emission ratio is given by 
${\delta=A\cdot\triangle V^\alpha}$, where $A$ is a constant, $\triangle V\approx V/(n+1)$ is the 
interstage voltage, and $\alpha$ is a coefficient which depends on the dynode material and geometric 
structure (typically $\alpha\approx$0.7-0.8).
For a voltage $V$ applied between the cathode and the anode, 
the gain is roughly $G\approx k\cdot V^{\alpha n}$, where $k$ is a constant. 
So the gain (or the PMT output signal amplitude) is proportional to the applied voltage $V$ and will 
increase as $V^{\alpha n}$ (in the linearity range of the PMT).

But with the applied high voltage the anode dark current will also increase (current in the 
PMT even when it is not illuminated). Major sources of dark current are 
thermoelectric emission of electrons from the materials, ionization of residual gases, glass 
scintillation, leakage current from imperfect insulation.
The resulting noise from the dark current is a critical factor in determining the low limit of 
light detection, in the optimization of the PMT gain ( via high voltage), especially when the 
rate of dark current change varies.

The choice of XP3462B PMT was made after studies of several other 3 inch and 3.5 inch photomultiplier tubes
on the matter of having good linearity, photocathode uniformity, high quantum 
efficiency, and good timing properties. Gain variations with HV and dark currents also were 
measured.   

In order to understand the limits imposed by the PMTs on the performance of the detector, 
several tests were performed on a set of candidate PMTs ~\cite{Amatuni96}.
All showed excellent linearity over a 3500:1 dynamic range of a reasonably chosen high 
voltage, as well as good time and amplitude resolutions. 
For pulses corresponding to photo-electron (pe) yield of ${N_{pe}=10^3}$, which is the expected 
signal from 1 GeV electron, the amplitude and time resolutions were ${\sigma_A/A \approx 4\%}$, 
and  ${\sigma_t\approx}$100-150 ps (measured with a pulsed variable intensity UV laser). 
These tests served as a guide for specifying requirements for the procurement of the PMTs.

\begin{figure}
\begin{center}
\epsfxsize=3.40in
\epsfysize=2.40in
\epsffile{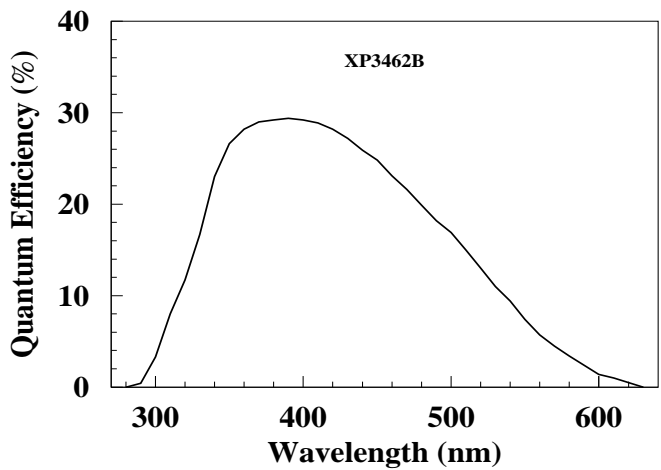}
\caption{\label{xp3462b-qe}
Typical quantum efficiency of the Photonis XP3462B PMT photocathode.}
\end{center}
\end{figure}

Following these tests, as a time and cost effective solution, Photonis XP3462B PMTs were chosen for
the equipment of the HMS and SOS calorimeters.
These 8-stage PMTs have a 3'' diameter ($\approx$68 mm) semitransparent bi-alkaline photocathode,
and a linear focused cube dynode structure with a peak quantum efficiency (QE) of
$\sim$29\% at 400\,nm (Fig.~\ref{xp3462b-qe}).
Using the criteria of high quantum efficiency, low dark current and high gain at relatively low HV,
the best PMTs (150 out of 180 available) were selected.
The negative operating voltages were set in the range $\sim$1.4-1.8 kV to match the gain 
$\sim 10^6$. The outputs were gain matched to within $\sim20\%$, and the 
remaining differences were corrected in software.

The PMT output signal may vary with respect to the incident photon's hit position on the
photocathode.
In general, this is caused by the photocathode and the multiplier (dynode section) non-uniformities.
Although the focusing electrodes of a phototube are designed so that electrons emitted from the 
photocathode are collected effectively by the first dynode, some electrons may deviate from their 
desired trajectories causing lower collection efficiency. The collection efficiency varies with 
position on the photocathode from which the photoelectrons are emitted and influences the spatial 
uniformity of a PMT. The spatial uniformity is also determined by the photocathode surface 
uniformity itself. If the cathode-to-first dynode voltage is low, the number of photoelectrons that 
enter the effective area of the first dynode becomes low, resulting in a slight decrease in the 
collection efficiency. 

For samples of PMTs the photocathode uniformity and effective diameter have been studied with 
a laser scanner. A $\sim$1 mm diameter fiber was positioned on the front of the PMT at a small 
distance from the photocathode. 
The light generated by the laser was split into two parts: one for the PMT scan, and another
to monitor incident light intensity by a photo-diode. 
The PMT was mounted on a special stand, which could be moved  remotely in 2-5 mm steps. 
At each position of the PMT, the coordinate information ($x_i$) from the scanner, PMT signal 
amplitude ($A^i_{pmt}$), and reference photo-diode signal ($A_0$) were readout and written to a data 
file. 
The PMT photocathode uniformity and effective diameter were found from the analysis of the
$A^i_{pmt}/A_0$ distribution versus $x_i$.
Nearly all the tested XP3462B PMTs had a photocathode of good uniformity and effective diameter of
no less than $\sim$2.8 inch.
The measured effective diameter only weakly depends on the PMT high voltage. 
This is likely an indication that the effective diameter is largely determined by the collection 
efficiency between the photocathode and the first dynode.

\begin{figure}
\begin{center}
\epsfxsize=3.40in
\epsfysize=2.40in
\epsffile{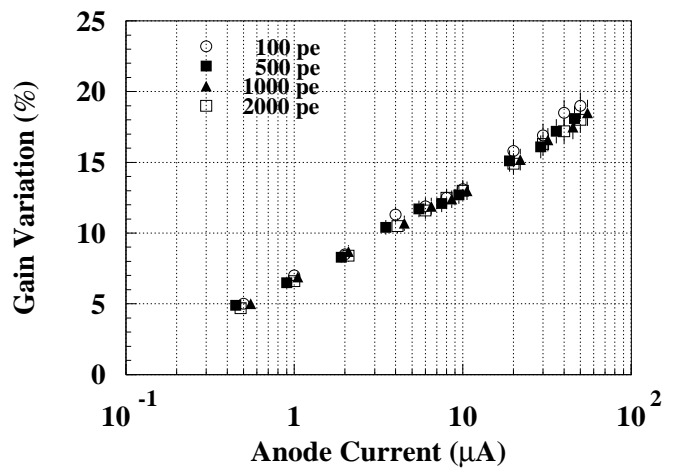}
\caption{\label{gain_var}
An example of the gain variation of a XP3462B PMT with mean anode current
measured at different light pulse heights, for the chosen HV distribution
among the base dynodes (for details see Ref.~\cite{Amatuni96}).
}
\end{center}
\end{figure}

Gain variation has been studied for the phototubes, under experimental conditions 
typical for CEBAF beam, as a function of the mean anode current, light pulse intensity and the 
high voltage distribution applied to the dynode system~\cite{Amatuni96}. These studies suggest
that at mean anode current $\sim 20~\mu$A the PMT gain may change up to $\sim 15\%$. 
Examples of gain variation with mean anode current measured at the different light 
pulse heights are shown in Fig.~\ref{gain_var}.

Samples of the assembled modules were tested in a magnetic field to evaluate the quality of the PMT
shields. At a fixed high voltage the blocks were illuminated through the ST connectors with a constant 
light intensity. Signal amplitude from the PMT was measured at gradually increasing values 
of the magnetic field. Measurements were performed at two different orientations of the PMT 
relative to the magnetic field: axial and transverse. 
As expected, the effect of the magnetic field was much stronger for the axial orientation. 
For both axial and transverse magnetic fields up to 2 Gauss, no effect was detected. Even at field 
values of about 4 Gauss, no effect was observed when the field was oriented transversely relative to 
PMT axis, while an axial field of the same strength reduced the PMT signal by 20--30\%.
We concluded that the PMT magnetic shields were sufficient, since in HMS and SOS detector huts
the calorimeters are located far from the magnets where fringe fields are less that 0.5 Gauss.

%---------------------------------------------------------------------------------------------------

\subsection{Studies on optical properties of TF-1 type lead glass blocks}
\label{tf1-block}
With its index of refraction $\sim$1.65, radiation length 2.74 cm and density of
$3.86~ {\rm g}/{\rm cm}^3$ TF-1 type lead glass is well suited for serving as \v{C}erenkov radiator
in electromagnetic calorimeters.
Note, the TF-1 radiative length found in different sources varies from 2.5 to 2.8\,cm. 
We cite the value obtained by means of PEGS4 (preprocessor for EGS4~\cite{EGS4}) and 
GEANT4~\cite{geant4} packages.
The fractional composition consists primarily of ${\rm PbO}$ (51.2\%), ${\rm SiO}_2$ (41.3\%), 
${\rm K}_2{\rm O}$ (3.5\%) and ${\rm Na}_2{\rm O}$ (3.5\%). 

Before assembly, the light transmittance of all the blocks was measured using
a spectrophotometer from the JLab Detector Group~\cite{Zorn}.
The wave-length was scanned from 200 nm to 700 nm in steps of 10 nm.
The blocks were oriented transversely, and the light intensity passing through the
10 cm thickness was measured.
Two measurements were carried out: with and without blocks
(to subtract dark current of the light detector and light loss in air).

Those measurements were repeated in 2008 on a set of blocks taken from the
decommissioned SOS calorimeter for re-use in the SHMS preshower counter.
The blocks had been in use under the beam conditions for 15 years, and thus
checks for possible degradation of the lead glass from radiation were
necessary.
Reliability of the measurements was checked by measuring spared, unused blocks and comparing with
1992 data.

\begin{figure}
\begin{center}
\epsfxsize=3.40in
\epsfysize=2.40in
\epsffile{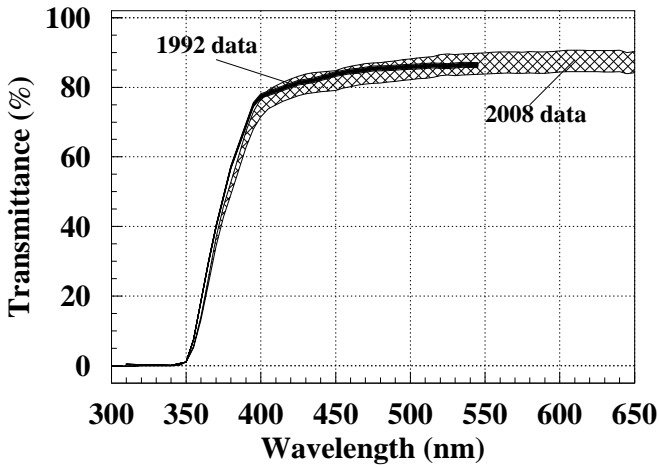}
\caption{\label{tf1_transp}
The light transmittance efficiency of TF-1 type lead-glass blocks.  The narrow dense band is data 
for unused blocks measured in 1992.  New measurements on used SOS blocks, carried out in 2008, are 
shown in the hatched band.}
\end{center}
\end{figure}

Results from 1992 and 2008 measurements are compared in Fig.~\ref{tf1_transp}.
Signs of marginal degradation can be noticed.

\begin{figure}
\begin{center}
\epsfxsize=3.40in
\epsfysize=2.40in
\epsffile{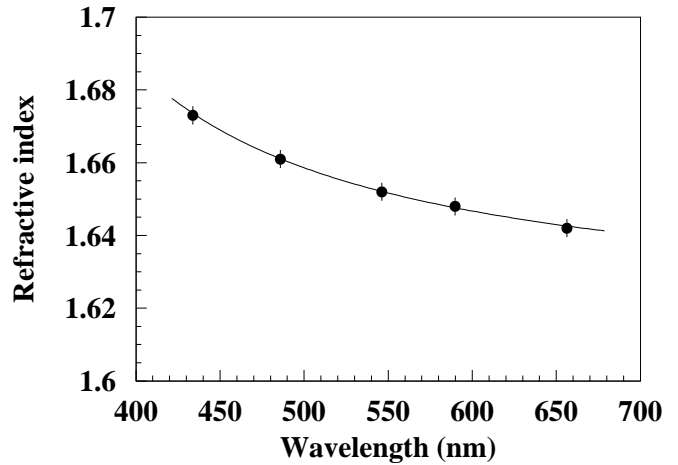}
\caption{\label{tf1_ri}
Refractive index of TF-1 lead-glass versus wavelength. The filled circle symbols are measurements, 
the curve is a fit to them in the form $n_{0} + \frac{n_{1}}{\lambda-\lambda_{0}},
n_{0}=1.617 \pm 0.004, n_{1}=10.4 \pm 2.3, \lambda_{0}=250.8 \pm 27.4$. 
}
\end{center}
\end{figure}

\begin{figure}
\begin{center}
\epsfxsize=3.40in
\epsfysize=2.40in
\epsffile{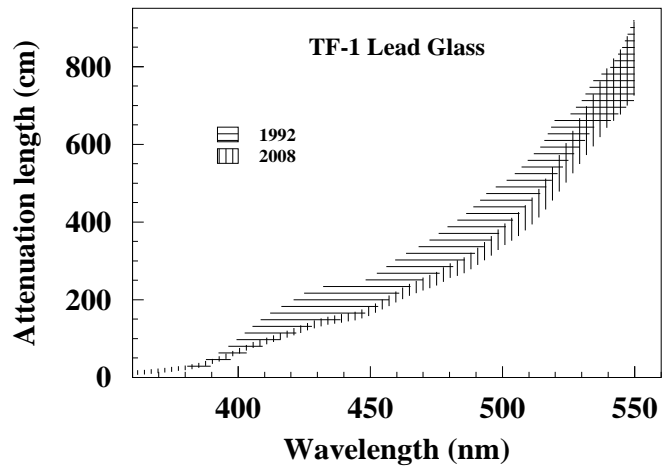}
\caption{\label{tf1_attl}
Light attenuation length of TF-1 lead-glass obtained from transmission measurements of the glass 
blocks in early 1990's (horizontally hatched area) and in 2008 (vertically hatched area). 
The hatched area for the 1990's is bounded by the best and the worst cases, while the area for 
2008 indicates the 2/3 majority of the cases.}
\end{center}
\end{figure}

In 1992 the transmittance of some of the blocks had been measured in the
longitudinal direction also.
From pairs of the transverse and longitudinal measurements
both refractive index (shown in Fig.\ref{tf1_ri}) and attenuation
length of the glass were extracted. From single measurements of the blocks in transverse
orientation, only attenuation lengths are extracted by assuming the nominal
refractive index of the glass of 1.65. As shown in Fig.\ref{tf1_attl},
the light attenuation length varies significantly in the range of sensitivity
of the XP3462B photocathode, and is $\sim$100 cm at the peak of sensitivity
$\sim$400 nm.
The slight shift between 1992 and 2008 year measurements is partly due to
different absolute calibrations of the setup. 

The block to block variations in light transmission were compensated by pairing high quantum 
efficiency PMTs with low transparency blocks and vice versa in the module
assemblies. Thus when all the PMTs were operated at a gain of $\sim 10^6$, the responses of 
modules to cosmic muons were equalized to within $\sim 20\%$.
For straight through muons, signal of 60-70 photo-electrons on average from a block,
and a pulse height resolution of $\sim 10-15\%$ were observed.

\begin{figure}
\begin{center}
\epsfxsize=3.40in
\epsfysize=2.40in
\epsffile{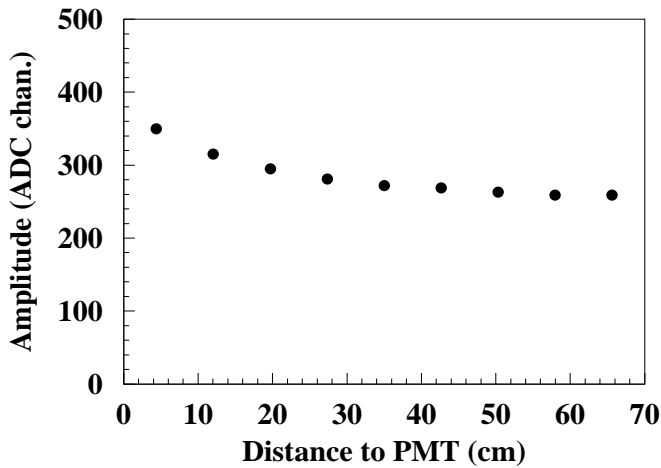}
\caption{\label{ydep}
Response of a prototype module to cosmic rays passing at different distances to the PMT. 
}
\end{center}
\end{figure}

The response of a module to cosmic rays passing at different distances to the PMT was studied. 
Two small ($5~{\rm cm}\times 5~{\rm cm}$) scintillator counters, placed on top and below
the module and aligned vertically, were used to localize particles and to trigger signal readout.
For single PMT modules, the signal variation at the edges was $\sim\pm15-20\%$
relative to the center (shown in Fig.~\ref{ydep}).
For the two PMT modules, variation of the summed signal was on the level of  
$\sim\pm7\%$. and the light output was about 1.5 times higher than for the
single tube case~\cite{Gasp92}.

\begin{figure}
\begin{center}
\epsfxsize=3.40in
\epsfysize=2.40in
\epsffile{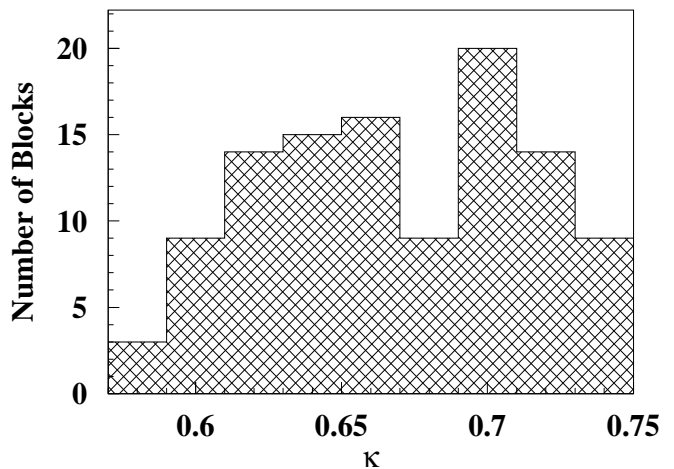}
\caption{\label{tf1_fom}
The distribution of lead-glass blocks by $\kappa = A_{\rm Blue}/A_{\rm Green}$.
See text for details. Blocks with $\kappa > 0.68$ were used in the SOS while
the rest were used in the HMS calorimeter.
}
\end{center}
\end{figure}

The relative light transmittance of all the assembled modules was measured by use of green 
and blue Light Emitting Diodes (LEDs). The ratio of light transmission efficiency for blue and 
green LEDs, $\kappa = A_{\rm Blue}/A_{\rm Green}$, (see Fig.~\ref{tf1_fom}) depends on optical 
properties of the blocks and is a measure of block quality. As the SOS spectrometer typically 
detected lower momentum particles than the HMS, the blocks with higher $\kappa$, and thus a higher 
transmission efficiency for \v{C}erenkov light, were used in the SOS calorimeter.
This also had the benefit of ensuring to some extent uniformity in the calorimeters.

Final equalization of the PMT output signals, determination of the function parameters for 
amplitude--distance corrections and overall calibration of the calorimeters were performed with 
electron beam, by using ``clean electron'' data after their installation in the spectrometer detector huts.

%---------------------------------------------------------------------------------------------------
\subsection{Choice of high voltage divider}
\label{xp3462-hvdiv}
\begin{figure}
\begin{center}
\epsfxsize=3.40in
\epsfysize=1.70in
\epsffile{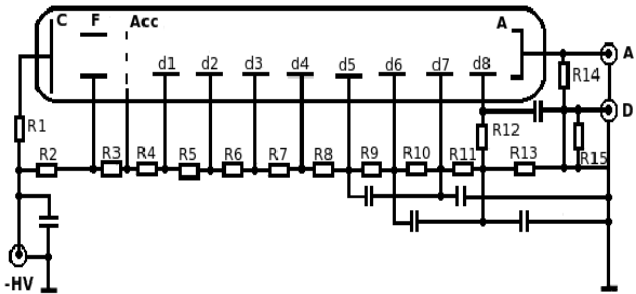}
\caption{\label{hv_divider}
Schematic of the high-voltage divider for XP3462B PMT. 
Selected with 1\%  tolerance and 1 W power resistors are:
$R1=10~M\Omega,~R2=5.62~k\Omega,~R3=33.2~k\Omega,~R4=110.4~k\Omega,
~R5=71.8~k\Omega,~R6=R7=R8=R9=47.5~k\Omega,~R10=59.3~k\Omega,~R11=95~k\Omega,~R12=50~\Omega,
~R13=71.8~k\Omega,~R14=R15=1.0~k\Omega$.
All capacitors are 10 nF.}
\end{center}
\end{figure}

Special studies were performed to optimize the PMT high voltage base design for the requirements of 
good linearity (better than 1\%), high rate capability and a weak variation of PMT gain with anode 
current~\cite{Amatuni96}.  
Two manufacturers~\cite{philips} recommended high voltage divider designs, optimized for high gain 
and linearity respectively. The bases had different relative fractions of the applied HV between 
the successive dynodes (from cathode to anode, including the focusing electrodes).
We selected a design, which is a compromise between the two,
but has also high anode current capability. 
This third design is a purely resistive, high current (2.3 mA at 1.5 kV), surface mounted divider 
($\sim 0.640~M\Omega$), operating at negative HV (see Fig.~\ref{hv_divider}). The relative fractions
of the applied HV between the  dynodes (from cathode to anode) are:
3.12/1.50/1.25/1.25/1.50/1.75/2.00/2.75/2.75. The supply voltage for a gain of $10^6$ is 
approximately 1750 V.

The PMT resistive base assembly is linear to within $\sim 2\%$ up to the peak anode current of 120 
$\mu$A ($\sim 5\times 10^4$ pe). The dark current is typically less than 3 nA.
The base has anode and dynode output signals.
Channel-to-channel adjustable high voltages are provided by a system of CAEN 
SY-403 high voltage power supplies (64 channel, $V_{max}$ = 3.0 kV, $I_{max}= 3.0~mA$).

%%%%%%%%%%%%%%%%%%%%%%%%%%%%%%%%%%%%%%%%%%%%%%%%%%%%%%%%%%%%%%%%%%%%%%%%%%%%%%%%%%%%%%%%%%%%%%%%%%%%

\section{Monte Carlo simulation codes}
\label{early_mc}

The first versions of simulation codes for the HMS/SOS calorimeters were based on the 
ELSS~\cite{ELSS} and EGS4~\cite{EGS4} packages for simulations of electromagnetic showers.
Dedicated code was added for \v{C}erenkov light generation, optical photon tracing and 
photoelectron knockout from PMT photocathodes.
The optics took into account light absorption in the lead glass, reflections from the block sides,
and passage through the optical coupling to the PMT photocathode.  However, the software did not 
take into account block to block variations of lead glass absorption length and electronic effects.
The first simulations revealed sufficient signal ($\sim$900 photoelectrons from a 1 GeV incident
electron), good linearity and reasonable resolution in the GeV range for the calorimeter designs.

Subsequent simulations of HMS calorimeter are based on the GEANT4 package, version 9.1.
The QGSP\_BERT physics list~\cite{qgsp-bert} was chosen to model hadron interactions, which is 
recommended by the GEANT4 developers for high energy physics calorimetry \cite{geant-www}.
This list includes the parton string model~\cite{parton-string} at energies above 12 GeV, 
intra-nuclear Bertini cascade~\cite{Bertini} below 9.9 GeV, and a nuclear 
evaporation model~\cite{nucl-evap} at low energies. The GHEISHA model~\cite{gheisha} is 
used at energies 9.5 -- 25 GeV. Electromagnetic processes are modeled to good accuracy 
within the framework of the GEANT4 standard electromagnetic package.

\begin{figure}
\begin{center}
\epsfxsize=3.40in
\epsfysize=2.40in
\epsffile{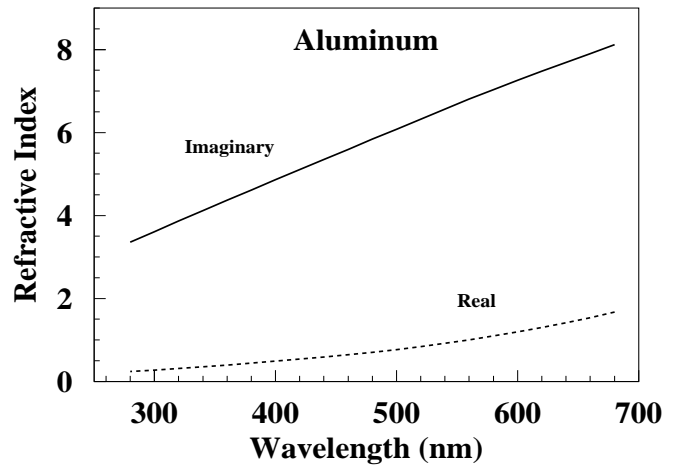}
\caption{\label{almylar}
Real (dashed line) and imaginary (solid line) refractive indexes of aluminum.}
\end{center}
\end{figure}

The code closely emulates the geometry and the composition of the detector. Particularly, the 
optical characteristics of the setup were thoroughly implemented in the light
tracing part of the code summarized below. The light attenuation length is 
randomly varied from block to block within the observed experimental limits 
(see Fig.~\ref{tf1_attl}).
The optical insulation of the module has multi-layer composition: air gap between aluminized Mylar 
and lead glass block, and Mylar support layer facing the block.  The reflective and absorptive 
properties of aluminum reflector are expressed by means of real and imaginary
parts of refractive index \cite{sopra} (see Fig.~\ref{almylar}).

Instead of GEANT4 optical photon handling,
the generated light is traced by means of a dedicated fast Fortran code which takes care of the 
modular construction and is suited to the particular geometry of the module.
Few compromises and simplifications took place in the code:
a strict rectangular geometry of the glass blocks is assumed;
all the boundaries are flat and perfectly smooth, diffuse reflections from the walls are neglected;
Rayleigh scattering in the glass is neglected as well;
nor the polarization of \v{C}erenkov light in the reflections/transmissions is taken into account.

Light reflectance from the block walls and passage from block to PMT photocathode is treated as
reflection/transmission from/through a plane-parallel plate sandwiched between two optical media
of different refractive indices. In the first case it is a layer of air in between the lead glass
and the reflector aluminum, in the second case it is a layer of optical grease between the lead glass
and PMT window glass. The layers are assumed thin enough to neglect light absorption,
and thick enough to neglect light interference effects.
With these assumptions, expressions for reflectivity and transmittivity of the boundaries were
derived in the limit of infinite series of Fresnel reflections/transitions
from the surfaces of the plate (similar to \cite{BornWolf}, p. 360).

This model was checked against GEANT4 calculations, and good agreement was found between the two.
In terms of the detector signal, the difference was less than a few percent.

A typical quantum efficiency of XP3462B photocathode (Fig.~\ref{xp3462b-qe}) is assigned to all the 
PMTs. Electronic effects are taken into account by assigning a random multiplicative ``gain'' 
factor to each channel in order to transform the number of photoelectrons into ADC channels. 
This factor is varied from channel to channel by 50\% around a mean value of 2. The electronic noise 
is modeled by adding a random pedestal of normal distribution with $\sigma=10$ ADC channels.
Both, the ``gain'' factor and the pedestal width roughly correspond to experimental conditions.

The projectiles are sampled at the focal plane of the spectrometer using the coordinate, angular and 
momentum distributions observed in the Meson Duality experiment \cite{mduality}.
The momenta are scaled to the settings of the studies.

Material traversed by particles before reaching the calorimeter smears the energy and coordinates 
of the particles. Therefore, all the material between the focal plane and calorimeter is also 
modeled (see Table~\ref{hms-mat}).

\begin{center}
\begin{table}
\caption{\label{hms-mat} Materials between HMS focal plane and calorimeter that are taken into 
account in the simulation. The listed positions are at the fronts of components}
{\centering  \begin{tabular}{|c|c|c|c|c|}
\hline
Component          & Material          & position   & thickness        & density  \\
                   &                   &  (cm)      &    (cm)          & $({\rm g}/{\rm cm}^3)$\\
\hline
DC2 gas            & Ethane/Ar         &  29.3      &  15              & 0.00143   \\
DC2 foils          &  Mylar            &            & 2$\times$0.00254 & 1.4       \\
S1X hodoscope      & BC408 scint.      &   77.8     &   1.067          & 1.032     \\
S1Y hodoscope      & BC408 scint.      &   97.5     &   1.067          & 1.032     \\
Aero. entrance     &   Al              &   40       &  0.15            & 2.6989    \\   
Aero. radiator     &  Aerogel          &            &      9           & 0.152     \\
Aerogel air gap    &    air            &            &    25.5          & 0.0012    \\
Aerogel exit       &    Al             &            &    0.1           & 2.6989    \\
Gas \v{C} gas      & $C_4F_1O$         &  198       &    150           & 0.0047    \\
Gas \v{C} wind.    &       Al          &            &  2$\times$0.1    & 2.6989    \\
Gas \v{C} mir.sup. &   Rohacell        &   230      &   1.8            & 0.050     \\
S2X hodoscope      &  BC408 scint.     &   298.8    &   1.067          & 1.032     \\
S2Y hodoscope      &  BC408 scint.     &  318.5     &  1.067           & 1.032     \\
Calo. support      &     Al            &   350      &    0.55          & 2.6989    \\
\hline
\end{tabular}\par}
\end{table}
\end{center}

%%%%%%%%%%%%%%%%%%%%%%%%%%%%%%%%%%%%%%%%%%%%%%%%%%%%%%%%%%%%%%%%%%%%%%%%%%%%%%%%%%%%%%%%%%%%%%%%%%%%

\section{Electronics and Calibration}
\label{electronic_calibr}

%---------------------------------------------------------------------------------------------------

\subsection{Electronics}
\label{calo-electronic}
\begin{figure}
\begin{center}
\epsfxsize=3.40in
\epsfysize=2.80in
\epsffile{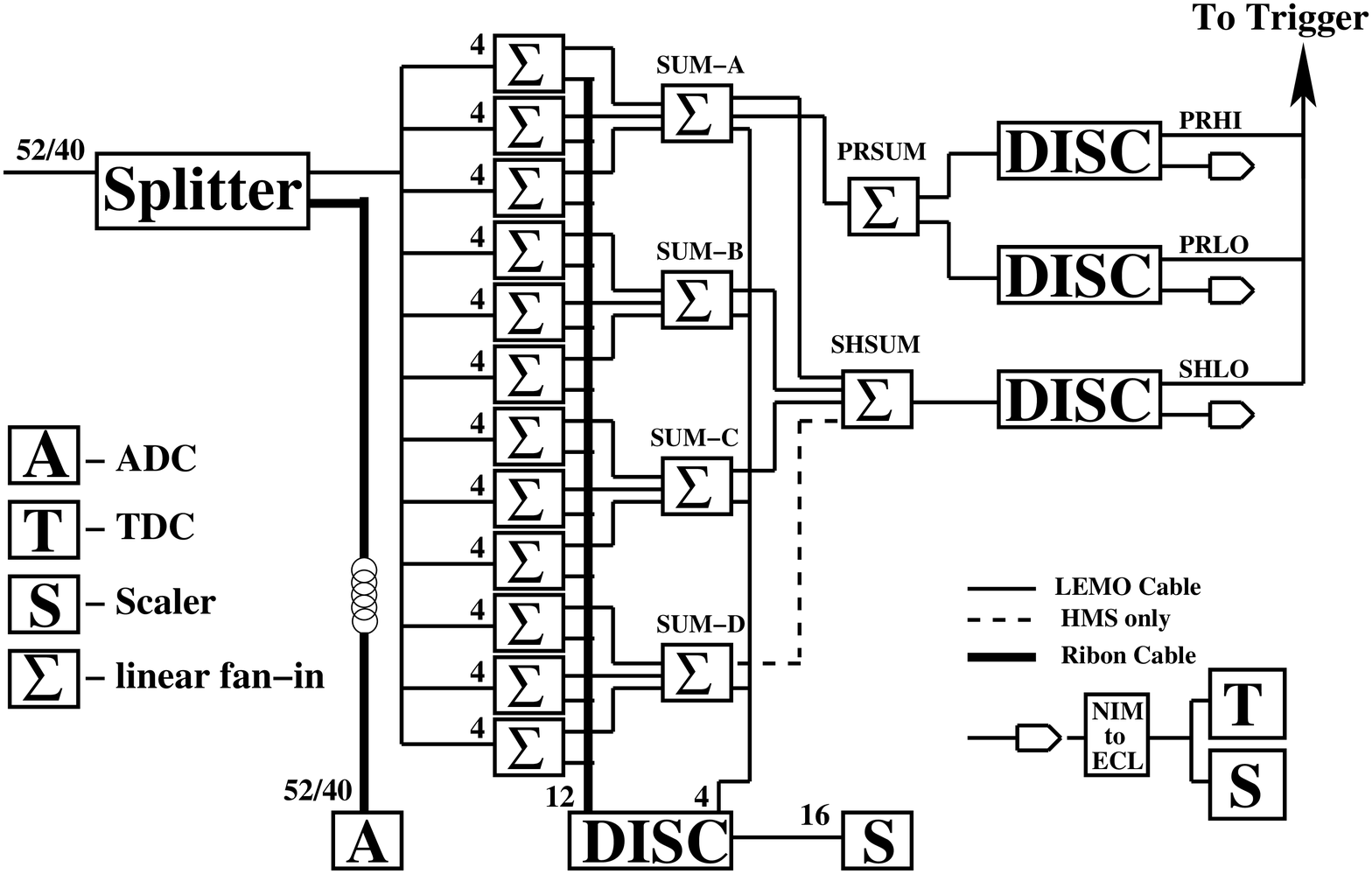}
\caption{\label{hcal_electronic}
Calorimeter electronics diagram. The numbers indicate the number of channels used in the HMS/SOS. 
For the SOS, the 4-th layer sum was not included in the trigger.
Adapted from \cite{arrington}.}
\end{center}
\end{figure}

The readout electronics were identical for both calorimeters.
The raw anode signals from the phototubes were taken from the detector hut to the electronics 
room through $\sim$30 feet RG58, then $\sim$450 feet RG8 coaxial cables. The signals were then 
split 50/50, with one output sent through 400 ns RG58 delay cable to a 64-channel LeCroy 1881M 
Fastbus ADC module, and the other to a Philips 740 linear fan-in modules to be summed. 
A schematic diagram of the electronics for the calorimeters is shown in Fig.~\ref{hcal_electronic}.

Data from the Fastbus modules were acquired in the ``sparsified'' mode, in which only significant data
were read from each ADC channel. The ADCs have programmable thresholds which were set ab initio fifteen
channels above zero. The zero (or ``pedestal'') of an ADC channel was determined at the beginning
of each run by creating 1000 artificial triggers. These thousand events show up as a narrow peak in a 
histogram of an ADC output. Typical pedestal widths were about 5-7  channels for the ADC gate width
$\sim$100 ns.
Then the new threshold for each ADC channel was calculated as three times the width above the
pedestal.
The automatically determined thresholds then can be used as input to the data
acquisition code such that it only reads out above the threshold, hence minimizing data flow.

Because of the high pion to electron ratio for some of the experiments, events are required to pass
loose particle identification cuts before generating a trigger. In order to have a high efficiency 
for electrons, a trigger was accepted as an electron if either the gas \v{C}erenkov  detector 
fired or if the electromagnetic calorimeter had a large enough signal.
The threshold on the gas \v{C}erenkov counter signal was typically set near the 1 pe level,
and the threshold on the calorimeter signal was set just above the pion peak, which is
independent of the spectrometer momentum setting.
This allowed for extremely high electron efficiency even if one of the two detectors had a low
efficiency.
On the other hand, the pion rejection was conditioned by the low, in this case, threshold on
the calorimeter signal.

Raw signals from the whole calorimeter and from the front layer alone are summed for use as an option
in the first level electronic trigger for $e/\pi$ discrimination~\cite{arrington}.
The fourth layer of the SOS calorimeter is not summed, since due to the 1.74 GeV maximum electron
energy in the SOS, most of the electromagnetic shower is contained in the first 3 layers, and
removing the last layer has almost no impact on the electron signal, but reduces the pion signal
(for straight through pions by 25\%).
The sum in the first layer (PRSUM) and the sum in the entire calorimeter (SHSUM) are discriminated 
to give three logic signals for the trigger: PRHI and PRLO are high and low threshold 
signals from the first layer, and SHLO from the entire calorimeter.
Also, groups of four modules are summed and sent through discriminators to scalers in order to call
attention to dead or noisy tubes.

The electron trigger (ELREAL) had two components: Electron High (ELHI) and Electron Low (ELLO).
ELLO was designed to trigger for all electrons, even those depositing low shower energy. Thus, it 
provided increased efficiency at the low electron momenta. ELLO required a \v{C}erenkov 
detector signal, a hodoscope signal (SCIN), and a shower signal (PRLO).
ELHI required a high calorimeter signal, but no \v{C}erenkov detector signal, and it was composed of 
preradiator high signal (PRHI), a three-out-of-four coincidence scintillator signal (SCIN)  and
the shower counter signal (SHLO).

%---------------------------------------------------------------------------------------------------

\subsection{Calorimeter Calibration}
\label{calo_calibr}

The ability of particle identification of a calorimeter is based on differences in the energy
deposition from different types of projectiles. The deposited energy is obtained by 
converting the recorded ADC channel value of each module into equivalent energy.
To obtain an accurate measurement of it, two main issues must be overcome: 
the light attenuation in the lead-glass block, and block to block PMT gain variation.

To correct the attenuation, the signal from each block is multiplied by a correction factor that 
depends on track position. This correction factor was different for the blocks with one and two PMT 
readouts. The correction was checked by looking at the distributions of corrected energy as a 
function of distance from the PMTs.

The PMT gains had been matched in the hardware in order to make the calorimeter trigger uniform
within acceptances of the calorimeters as much as possible.
At first, using scattered electrons in each spectrometer the operating high voltages for the 
PMTs were adjusted so that the ADC signals were nearly identical (to $\sim 10\%$) for 
blocks in the same layer.
Electrons with larger momenta are bent less in the spectrometer, and populate the bottom 
blocks in the calorimeter. Because the bottom blocks detect higher energy electrons, their gain 
must be kept lower than for the top blocks so that the output signals are of the same size. 
Therefore, setting the gain such that the output signal is constant as a function of vertical 
position in the calorimeter means having a gain variation between the blocks roughly equal to the 
momentum acceptance of the spectrometers ($\sim 20\%$ in the HMS, $\sim 40\%$ in the SOS). 
The output signals were made equal (rather than gains) in order to make the calorimeter trigger 
efficiency as uniform as possible over the entire calorimeter.

The data analysis procedure corrects for the gain differences in the process of calorimeter
calibration. Good electron events are selected by means of gas \v{C}erenkov detector.
The standard calibration algorithm~\cite{amatuni} is based on minimization of the variance
of the estimated energy with respect to the calibration constants, subject to the constraint that
the estimate is unbiased (relative to the primary energy).
The momentum of the primary electron is obtained from the tracking in the magnetic field of the
spectrometer.

The deposited energy per channel is estimated by
\begin{equation}
\label{eq:calo-module-e}
{ e_i = c_i \times (A_i - ped_i) \times f(y)  },
\end{equation}
where $i$ is the channel number, $c_i$ is the calibration constant, $A_i$ is the raw ADC signal,
$ped_i$ is the pedestal position, $f(y)$ is correction for the light attenuation for the
horizontal hit coordinate $y$.

Due to the segmentation in the vertical direction, the calorimeters have a coarse tracking 
capability which is helpful when separating multiple tracks 
(see, for instance, \cite{pruning}), \cite{tvaskis}).  In the calorimeter analysis code
hits on adjacent blocks are grouped into clusters for which the deposited energy and
center of gravity are calculated.  These clusters are matched with tracks from the 
upstream detectors if the distance from the track to cluster in the vertical direction is less 
than a predefined ``slop'' parameter (usually 7.5~cm).

\begin{figure}
\begin{center}
\epsfxsize=3.40in
\epsfysize=2.40in
\epsffile{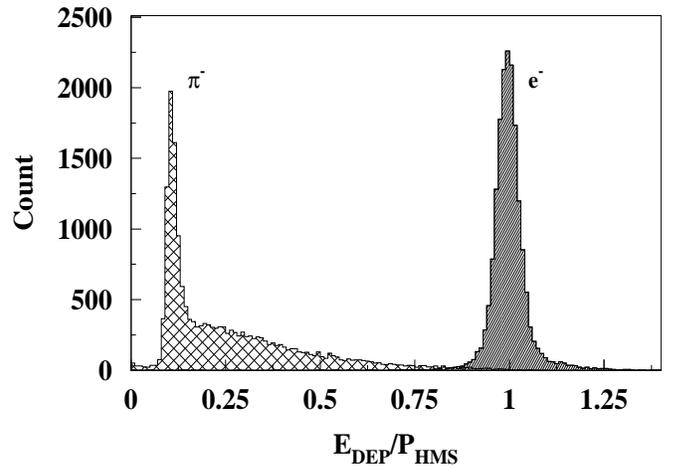}
\caption{\label{fpi2_edep}
Energy deposition in the HMS calorimeter from 3 GeV/c pions (hatched histogram) and electrons 
(filled histogram) in E01-004 (Fpi-2) experiment.}
\end{center}
\end{figure}

The calorimeter energy corresponding to a track is divided by the track momentum and used for 
particle identification. In the few GeV/c range pions and electrons are well separated
(see Fig.~\ref{fpi2_edep}), a cut at 0.7 ensures an electron detection efficiency better than 
99\% and ~30:1 pion suppression (see \cite{malace} and Section~\ref{hcal_scal_perform}).

%%%%%%%%%%%%%%%%%%%%%%%%%%%%%%%%%%%%%%%%%%%%%%%%%%%%%%%%%%%%%%%%%%%%%%%%%%%%%%%%%%%%%%%%%%%%%%%%%%%%
\section{Performance of HMS/SOS calorimeters}
\label{hcal_scal_perform}

\subsection{Selection of calorimeter experimental data}
\label{exp-data}

For these studies, HMS calorimeter data from the E01-004 (Fpi-2) \cite{fpi2} and E00-108
(Meson Duality) \cite{mduality} experiments have been collected for comparison with simulations.
Fpi-2 measured the charged pion form factor at $Q^2$=1.6 and 2.45 (GeV/c)$^2$ via exclusive pion 
production, while Meson Duality looked for signatures of quark-hadron duality in semi-inclusive 
pion production. The experiments ran back to back in summer of 2003, and both detected pions in the 
HMS in coincidence with electrons in the SOS. Fpi-2 also detected electrons at elastic 
kinematics for HMS acceptance studies.
Some other JLab experiments also used HMS or SOS calorimeters for good pion rejection and studied
the devices, such as E89-008 \cite{arrington}, E02-019 \cite{fomin}, E03-103 (\cite{seely},
\cite{daniel}).

In order to obtain high purity samples of electrons and pions, tight cuts were applied to 
spectrometer events.  Only events with single tracks in HMS and SOS passing through the collimators
were used. Spectrometer acceptances were restricted to ensure good tracking accuracies, and,
on the HMS side, efficient particle identification with gas \v{C}erenkov counter.
Electrons in HMS were identified by applying a high cut on the gas \v{C}erenkov signal
greater than 4 photoelectrons, while pions were identified with null signal.

Pion samples were selected in the HMS from $(e'\pi)$ coincidence events, by posing tight electron PID 
cuts on the SOS gas \v{C}erenkov detector signal greater than 3 photoelectrons, and normalized
energy deposition in the SOS calorimeter $E_{Dep}/P_{SOS}$ greater than 0.9.
Furthermore, a coincidence timing cut $\mid$cointime$\mid < $1~ns was also applied.
Accidental events were selected and subtracted from energy deposition histograms by off coincidence
timing cut 3~$< \mid$cointime$\mid < $~13~ns.

In addition, a kinematic cut of exclusive pion production on the missing mass was applied for Fpi-2.
Finally, for these studies HMS calorimeter was calibrated on a run by run basis.
Examples of the resultant distributions of the energy depositions in the calorimeter from incident
electrons and pions are shown in Fig.~\ref{fpi2_edep}.

%---------------------------------------------------------------------------------------------------

\subsection{Resolution of HMS/SOS calorimeters}
\label{hcal-scal-res}

\begin{figure}
\begin{center}
\epsfxsize=3.40in
\epsfysize=2.40in
\epsffile{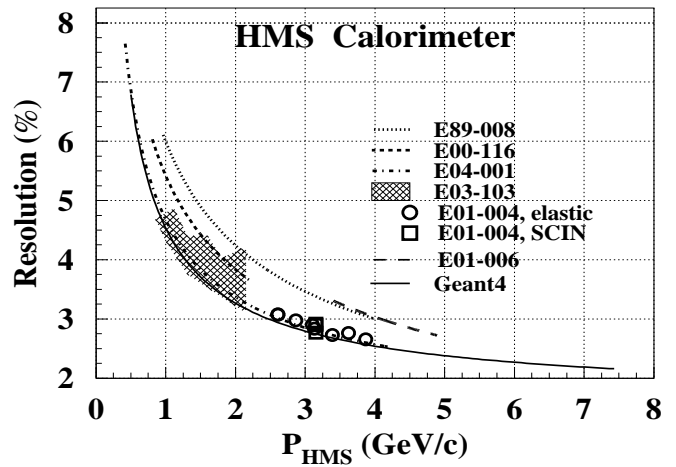}
\caption{\label{hcal_sigma}
Resolution of the HMS calorimeter. Dotted, dashed, dash-dotted and long-dashed lines are fits to data
taken from E89-008~\cite{arrington,niculescu}, E00-116~\cite{malace}, E04-001~\cite{fpi2} experiments,
and from online analysis of E01-006~\cite{RSS} experiment respectively.
The dashed area represents scattered data from E03-103~\cite{seely} experiment. 
The empty symbols are our re-analysis of the E04-001 experiment. 
The solid line is a fit to the GEANT4 calculations (see text for details).}
\end{center}
\end{figure}

We define calorimeter resolution as the width of a Gaussian fit to the electron 
peak (Fig.~\ref{fpi2_edep}) in the distribution of energy deposition.

The resolution of HMS calorimeter from a number of Hall C experiments is compared with simulation in 
Fig.~\ref{hcal_sigma}. Experiments before the modification of the detector in 1998
(see subsection~\ref{hcal_module}), like E89-008 shown in the figure, 
report resolution $\sim6\%/\sqrt{E}$ ($E$ in GeV) (\cite{arrington}, \cite{niculescu}).
Experiments carried out afterward found improved energy resolution. 
Exception is the E99-118 experiment~\cite{tvaskis} with resolution 8\%/$\sqrt{E}$.
E00-116, the first experiment to actually analyze data 
with the modified calorimeter, states resolution $5.4\%/\sqrt{E}$ \cite{malace}.
E03-103, despite of a gain shift problem in the calorimeter electronics, obtained somewhat
scattered data close to the simulation \cite{seely}.
E04-001 got very good resolution in the wide range of HMS momenta, in agreement with 
simulation, presumably due to relatively low rate, good tracking conditions, and run by run 
calibration \cite{mamyan}.
A somewhat worse resolution is obtained from online analysis of the
E01-006~\cite{RSS} experiment at high energies up to $\sim$4.7 GeV/c.

As for the SOS calorimeter, an on-line data analysis during the Hall C Spring03 experiments
(E00-002, E01-002, E00-116) gave a resolution of $6\%/\sqrt{E}+1\%$,
within the range of SOS momentum setting 0.5 -- 1.74 GeV/c.
The E00-108 experiment reported a resolution consistent with $\sim5\%/\sqrt{E}$ for SOS momentum
range 1.2 -- 1.7 GeV/c \cite{navasardyan}.

In general, resolution from an experiment depends on multiple factors related both to hardware and
software. Some of them, like trigger rate, background rate, performance of tracking detectors, 
tracking algorithm itself affect performance of the calorimeter indirectly, through the tracking
conditions. 
Other factors, like electronic noise, stability of high voltage supply, low energy background,
calibration affect the performance directly.

The conventional 3-parameter fit \cite{pdg} to the simulated HMS resolution (in \%) gives
a dependence on energy in the form $3.75/\sqrt(E) \oplus 1.64 \oplus 1.96/E$. The first term
is purely of stochastic origin, the second term reflects systematics from non-uniformity of the
detector and calibration uncertainty, the third term, poorly constrained here by limited statistics,
comes from electronic noise.
In the simulated data stochastic and systematic terms dominate, electronic noise is tangible only
at low energies $\lesssim$1.5 GeV. 

Overall, the resolution of the HMS/SOS calorimeters is close to resolutions of the lead-glass 
calorimeters of similar thicknesses (see \cite{Avakian98} and references therein, 
also \cite{e705_93,wa91_95}).

%---------------------------------------------------------------------------------------------------
\subsection{Electron detection efficiency and pion rejection}
\label{hcal-eff-rej}

\begin{figure}
\begin{center}
\epsfxsize=3.40in
\epsfysize=3.40in
\epsffile{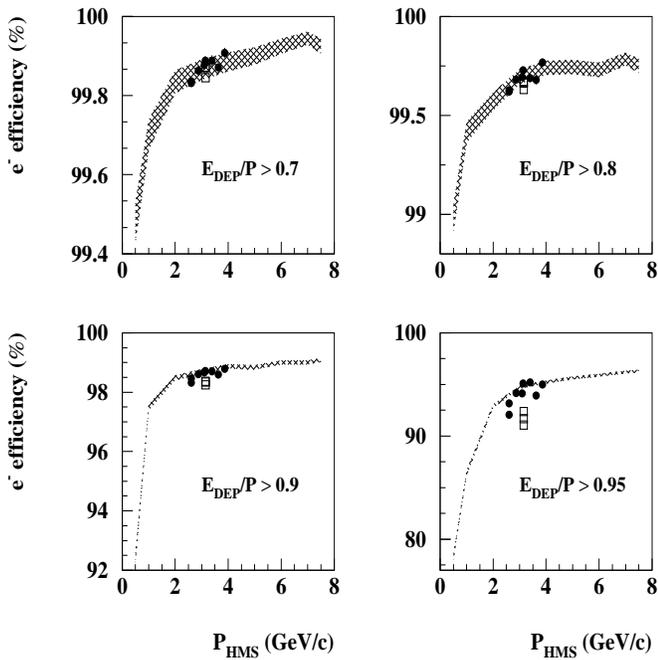}
\caption{\label{hms_eff_vs_p}
Efficiency of electron detection in HMS calorimeter at different momenta and for different cuts
on the normalized energy deposition. The shaded areas represent results from GEANT4 simulations.
The solid circles and empty boxes are data from Fpi-2 experiment taken at elastic scattering
kinematics and from exclusive pion production respectively.
}
\end{center}
\end{figure}

\begin{figure}
\begin{center}
\epsfxsize=3.40in
\epsfysize=2.40in
\epsffile{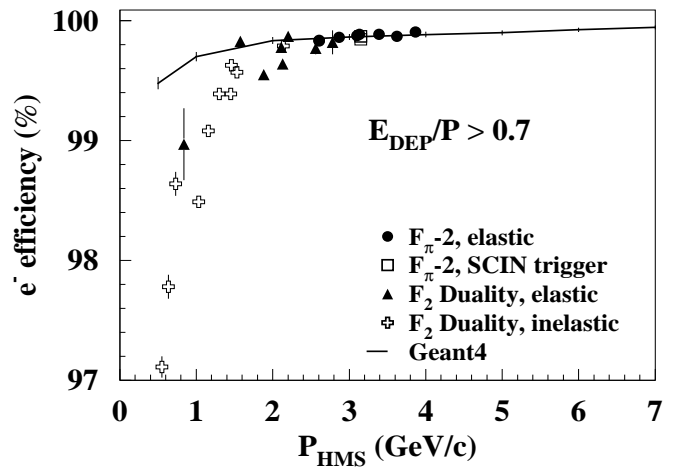}
\caption{\label{hms_cut_eff_07}
Efficiency of electron detection in HMS calorimeter versus momentum of the spectrometer, 
with cut on the normalized energy deposition at 0.7. GEANT4 simulation is compared to the 
Fpi-2 reanalysis and data from the inclusive resonance electroproduction 
experiment~\cite{niculescu}.
}
\end{center}
\end{figure}

The experimental efficiency of electron detection, which is defined as the fraction of
events with the normalized energy
deposition above threshold, at momenta within the range 2.8--4.1 GeV/c, for different
cuts is in reasonable agreement with the simulation (see Fig.~\ref{hms_eff_vs_p}).
The simulation predicts a steady rise of $e^-$ detection efficiency with energy due to the
improvement in resolution. However, as shown in Fig.~\ref{hms_cut_eff_07} there is a growing
disagreement with experiment for energies below 2~GeV.

\begin{figure}
\begin{center}
\epsfxsize=3.40in
\epsfysize=3.40in
\epsffile{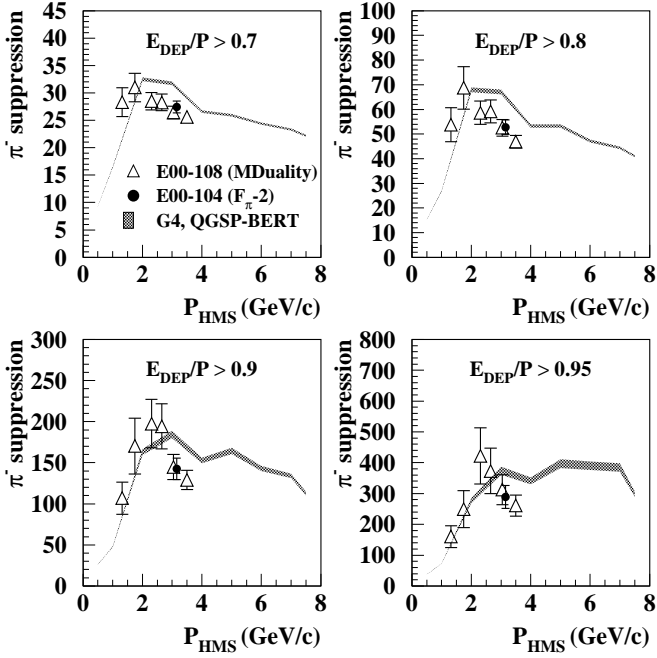}
\caption{\label{hms_cut_sup_vs_p}
Pion suppression factor of HMS calorimeter versus momentum of the spectrometer at different cuts
on the normalized energy deposition. The GEANT4 simulation (shaded area) is compared to data from 
Meson Duality and Fpi-2 experiments. A cut on the missing mass is added for the selection of
exclusive pion production in Fpi-2.}
\end{center}
\end{figure}

The $\pi^-$ suppression factor, the ratio of total number of pionic events and misidentified as 
pions, at different momenta and cut values is shown in Fig.~\ref{hms_cut_sup_vs_p}.
Experimental data for comparison are mostly from the Meson Duality experiment. 
At 3 GeV/c there are data from Fpi-2 as well. The Fpi-2 data are presumably of better quality 
due to favorable background conditions and exclusive kinematics for pion production. 
Good agreement between the two experiments at 3 GeV ensure the quality of the pion suppression data
found in Meson Duality.

Both experiment and simulation show a momentum dependence of the suppression factor peaking at
several GeV/c. 
While in the experiment the peak value is reached at $\sim$2.5 GeV/c independent of the cut,
in the Monte Carlo it shifts to higher momenta as the cut is raised.
Overall, agreement between experiment and simulation is satisfactory for the rejection studies.

Few Hall C experiments report on the rejection capabilities of HMS or SOS calorimeters. E89-008, 
the first Hall C experiment \cite{arrington} states pion suppression by 25:1 for $E_{Dep}/P> 0.7$ 
at 1 GeV/c HMS momentum, which agrees with experimental data in this study,
and fast improvement with energy due to moving the threshold to higher positions.
E94-014 reports a pion rejection 95\% at SOS 
momenta 1.4-1.5 GeV/c for the cut value of 0.7 \cite{armstrong}.
Note that these two experiments ran before the calorimeters had been modified.
The same rejection is reported in E00-108 for HMS at 1.7 GeV/c \cite{navasardyan}, again in rough
agreement with this study.
The higher suppression factor obtained in this study comes from the cleaner selection
of the particle samples (see subsection~\ref{exp-data}).

\begin{figure}
\begin{center}
\epsfxsize=3.40in
\epsfysize=2.40in
\epsffile{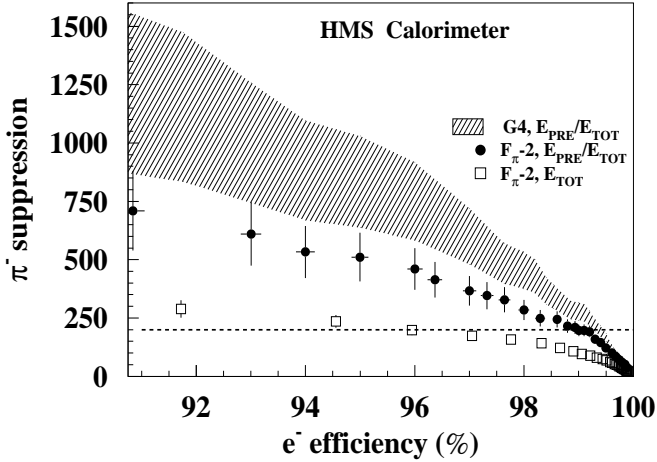}
\caption{\label{hms_sup_vs_ef}
$\pi^-$ suppression factor versus $e^-$ detection efficiency from 2-dimensional 
Preshower/Shower separation (closed symbols) and ordinary 1-dimensional rejection (open symbols)
in HMS calorimeter, obtained from analysis of Fpi-2~\cite{horn} experimental data at 3.153 GeV/c
HMS setting. Here, the forward layer of the detector served as the Preshower.
The shaded area represent results from GEANT4 simulation on the 2-dimensional separation at
3 GeV/c.
}
\end{center}
\end{figure}

Segmentation of HMS calorimeter allows for using the difference in longitudinal development of
electromagnetic and hadronic showers for PID. In particular, energy deposition in the forward
layer is most indicative. This is elaborated in subsection \ref{shms-calo-perf},
with regard to the SHMS calorimeter.
As it is seen in Fig.~\ref{hms_sup_vs_ef}, one can gain substantially in PID capability
of the HMS counter, by combining energy depositions in the first layer and in the whole calorimeter.
Relative to the ordinary rejection, at the 3 GeV/c spectrometer setting, improvement in pion
suppression is more than twice at low electron detection efficiencies 90 -- 95\%,
and $\sim$1.5 times at high efficiencies above 99.7\%.
Alternatively, one can keep the suppression factor constant and gain in detection efficiency. 
For instance, at 250:1 $\pi^-$ suppression factor one can boost $e^-$ detection efficiency from 
$\sim$93\% to $\sim$98.5\%.

%---------------------------------------------------------------------------------------------------
\subsection{Long-term stability of calorimeters}
\label{hcal-scal-stab}
The HMS/SOS calorimeters' resolution shows only slight changes during the years of usage
(see Fig.\ref{hcal_sigma}).
These changes include variations in electronics, 
calibration technique and possible degradation of the calorimeter components.

Stability of both calorimeters also has been evaluated by tracking changes in the ADC pedestal
and PMT gain values. These values have been found to be stable within accuracy of the measurements 
during the entire time of operation. 
The long-term stability of the calorimeters' responses have been monitored by tracking the variations 
in the width of normalized energy deposition (E/p) distribution, and variations in the PMT gain 
calibration constants from run to run, and from experiment to experiment. 
No significant degradation of HMS/SOS calorimeters' performances
after 15 years of operation have been noticed.

%%%%%%%%%%%%%%%%%%%%%%%%%%%%%%%%%%%%%%%%%%%%%%%%%%%%%%%%%%%%%%%%%%%%%%%%%%%%%%%%%%%%%%%%%%%%%%%%%%%%

\section{SHMS Calorimeter}
\label{shms_calo}

%---------------------------------------------------------------------------------------------------
\subsection{Design construction}
\label{shms-calo-constr}

As a full absorption detector, the SHMS calorimeter is situated at the very end of detector stack of 
the spectrometer~\cite{CDR-12}.
The relatively large beam envelope of the SHMS dictated a different calorimeter design from HMS/SOS,
with a wider acceptance coverage. In order to exclude possible energy leaks at higher energies,
it was necessary to consider a shower counter for SHMS thicker than in HMS.
The deeper calorimeter, the less energy leak of the electromagnetic shower from the radiator,
but more light loss due to absorption in the glass and reflections is expected.
Therefore, there should be an optimum in the detector dimension along the particle trajectory.
For an energy range of a few tens of GeV it was found that the optimum is at the radiator length of
$\sim$40 cm~\cite{Avakian96,Binon}.

The general requirements for the SHMS calorimeter are: \\
- Effective area: $120 \times 140~{\rm cm}^2$;  \\
- Total thickness: $\sim$20 rad. length;  \\
- Dynamic range: 1.0 - 11.0 GeV/c;        \\
- Energy resolution: $\sim 6\%/\sqrt E $, $E$ in GeV; \\
- Pion rejection: $\sim$100:1 at $P\gtrsim$1.5-2.0 GeV/c;  \\
- Electron detection efficiency: $>98\%$.

%---------------------------------------------------------------------------------------------------
\subsection{Studies of different versions and choice of assembling}
\label{choice-assembl}
 
A few different versions of calorimeter assembly for the SHMS spectrometer have been considered 
(\cite{A-Mkrt06,A-Mkrt07,H-Mkrt10}) before it was optimized for cost/performance.
A possible choice is a construction similar to the HMS and SOS calorimeters.
An alternative is a calorimeter similar to HERMES~\cite{Avakian98} and Hall A~\cite{Alcorn} 
shower counters. The goal of these studies was to explore a few proposed versions of the SHMS 
calorimeter based on commercially produced lead glass.

The configurations considered are a total absorption part (called Shower in the following), or a 
combination preshower and shower parts (``Preshower+Shower'' in the following). 
The Preshower is a slab of a few radiative length thick lead-glass before the Shower part. For each 
version the energy resolution, electron detection efficiency and pion/electron separation 
capabilities were determined by simulations.
The Shower and Preshower were made from modules, which consist of an optically 
isolated rectangular lead-glass block and optically coupled to it a PMT.
 
For all versions we assumed only modular construction of the calorimeters since this gives more 
flexibility in assembling and allows for localizing the position of energy deposition clusters. 
Different types and sizes of the lead-glass blocks were also considered. We found the energy 
resolution for all versions with and without Preshower to be nearly similar, but different  versions 
required different number of modules (channels) to cover the acceptance of the SHMS.
Adding a Preshower dramatically improves the $\pi/e$ rejection factor.

\begin{figure}
\begin{center}
\epsfxsize=3.40in
\epsfysize=3.40in
\epsffile{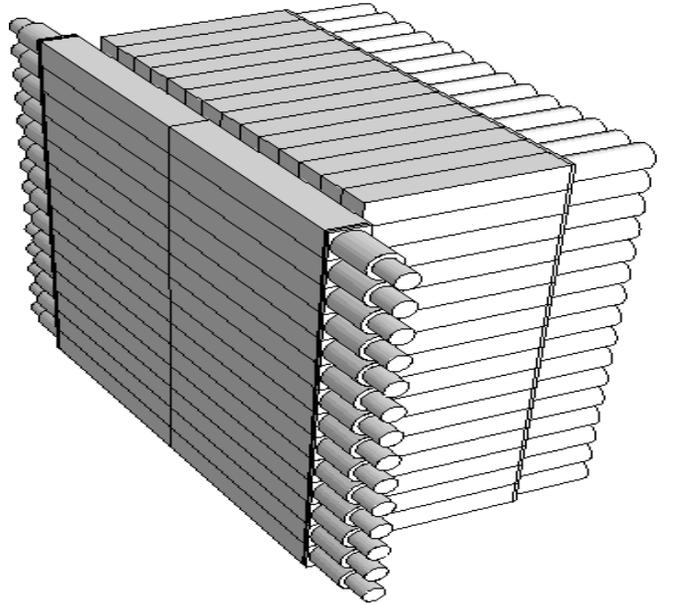}
\caption{\label{shms_calo_sk} A sketch of SHMS calorimeter. Shown are Preshower (on the left) and 
Shower parts. Support structures are omitted.  
}
\end{center}
\end{figure}

Our studies allowed selection of the optimum calorimeter geometry while maintaining the good energy 
resolution and pion rejection capabilities.
The newly designed SHMS calorimeter consists of two parts (see Fig.~\ref{shms_calo_sk}):
the main part at the rear (Shower), and Preshower before the Shower to augment PID capability of the
detector.
An optimal and cost-effective choice was found by using available modules from 
HERMES calorimeter for Shower part, and modules from SOS calorimeter for Preshower.
With this choice the Shower becomes 18.2 radiative length deep and almost entirely absorbs showers
from $\sim$10 GeV electromagnetic projectiles, and Preshower becomes 3.6 radiation length thick.

%---------------------------------------------------------------------------------------------------

\subsection{Description of constructive elements}
\label{shms-calo-elements}

The SHMS Preshower radiator consists of a layer of 28 TF-1 type lead glass blocks
from the calorimeter of the retired SOS spectrometer in Hall C, stacked in two columns
in an aluminum enclosure (not shown in Fig.~\ref{shms_calo_sk}).
28 PMT assemblies, one per block, are attached to the left and right sides of the enclosure.
The Shower part consists of 224 modules from the decommissioned HERMES detector~\cite{Avakian98}
stacked in a ``fly eye'' configuration of 14 columns and 16 rows.
$\sim 120\times130~{\rm cm}^2$ of effective area of detector covers the beam envelope at the
calorimeter.

The Preshower enclosure adds little to the material on the pass of particles. On the front and back
are 2'' Honeycomb plate and a 1~$mm$ sheet of aluminum respectively, which add up to 1.7\% of
radiation length only.
The optical insulation of the $10~{\rm cm}\times 10~{\rm cm}\times 70~{\rm cm}$ TF-1 blocks
(see Section~\ref{hcal_scal} for details) in the Preshower is optimized to minimize
the dead material between them, without compromising the light tightness.
First, the blocks are loosely wrapped in a single layer of 50~$\mu$m thick reflective aluminized
Mylar film, with Mylar layer facing the block surface.
Then, every other block is wrapped with a 10~$cm$ wide strip of 50~$\mu$m thick black Tedlar film,
to cover its top, bottom, left and right sides but the circular openings for the PMT attachments.
Looking at the face of detector, the wrapped and unwrapped blocks are arranged in a chess pattern.
Insulation of the remaining front and back sides of the blocks are provided by facing inner surfaces
of the front and rear plates of the enclosure, covered also with Tedlar.
In addition, a layer of Tedlar separates the left and the right columns.

The PMT assembly tubings are screwed in 90~$mm$ circular openings on both sides
of the enclosure.
The spacing of the openings matches the height of the blocks, so that a PMT faces to each of
the blocks.
The 3'' XP3462B PMTs are optically coupled to the blocks using ND-703 type Bycron grease of
refractive index 1.46.

The HERMES modules to be used in the Shower part are similar in construction to the HMS/SOS modules
but differ in details. The radiator is an optically isolated $8.9\times8.9\times50~{\rm cm}^3$
block of F-101 lead-glass, which is similar to TF-1 in physical parameters.
The typical density of F-101 type lead-glass is 3.86 ${\rm g}/{\rm cm}^3$, radiation length 2.78 cm, 
and refraction index ~1.65. 
The chemical composition of F-101 is: $Pb_2O_4$ (51.23\%), $SiO_2$ (41.53\%), $K_2O$ (7\%) and 
$CeO$ (0.2\%) by  weight~\cite{Avakian96}. 
The small amount of Cerium, added for the sake of radiation hardness
(\cite{Kobayashi}, \cite{Adams}), absorbs light at small wavelengths, and thus restricts the band of
optical transparency to higher wavelengths (see Fig.~\ref{f101_tf1_attl}).

\begin{figure}
\begin{center}
\epsfxsize=3.40in
\epsfysize=2.40in
\epsffile{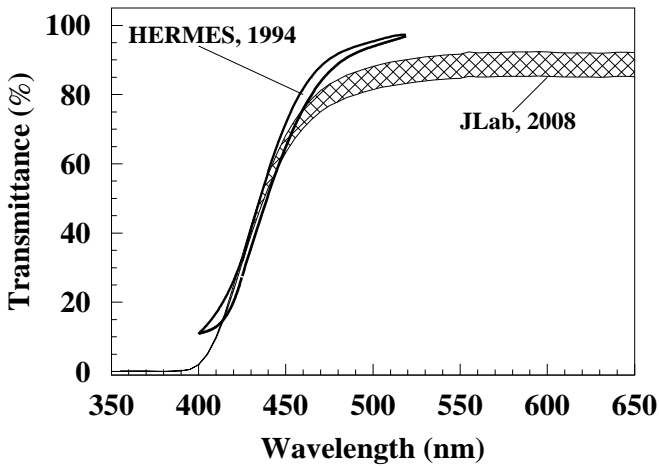}
\caption{\label{f101_norad_trans}
Transmittance of unused (not radiated) F-101 lead-glass blocks. The hatched area 
represents results from JLab 2008 year measurements, narrow band within solid lines 
is 1994 year data from HERMES collaboration \cite{Avakian96}.
}
\end{center}
\end{figure}

Results of F-101 type lead-glass block transmittance measurements are shown in
Fig.~\ref{f101_norad_trans}.
For unused blocks, a $\sim 10\%$ shift in transmittance has been found between the 1994 year
measurements by the HERMES collaboration~\cite{Avakian96} and our measurements at JLab in 2008.
We believe that the shift between the two sets of 
measurements is due to different calibration techniques of the setups. 

\begin{figure}
\begin{center}
\epsfxsize=3.40in
\epsfysize=2.40in
\epsffile{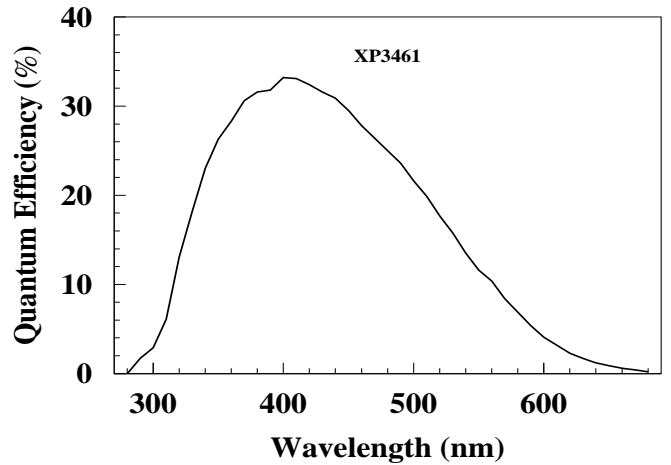}
\caption{\label{xp3461-qe} Typical quantum efficiency of a photocathode of the XP3461 PMT used
in the simulation, derived from the typical radiant sensitivity from the vendor.  
}
\end{center}
\end{figure}

Each F-101 block is coupled to a 3'' XP3461 PMT from Photonis, with green extended bialkali 
photocathode, of the same sizes and internal structure as the XP3462B in the HMS/SOS calorimeters
and in the Preshower.
Typical quantum efficiency of the photocathode is $\sim30\%$ for $\lambda\sim$400 $nm$ 
light (see Fig.~\ref{xp3461-qe}), and the gain is $\sim10^6$ at $\sim$1500 V.
Silgard-184 silicone glue of refractive index 1.41 is used for optical coupling of the PMTs to 
lead-glass blocks.

A $\mu$-metal sheet of 1.5 mm thickness and two layers of Teflon foil are used for magnetic
shielding and electrical insulation of the PMTs.
The blocks are wrapped with 50 $\mu$m aluminized Mylar and 125 $\mu$m black Tedlar paper for
optical insulation.
A surrounding aluminum tube which houses the $\mu$-metal, is fixed to a 
flange, which is glued to the surface of the lead-glass. The flange is made of titanium, which 
matches the thermal expansion coefficient of F-101 lead-glass~\cite{Avakian96}.

Beyond simple repairs, no adjustment has been made to the original HERMES construction of the modules
for re-use in the SHMS calorimeter.

%---------------------------------------------------------------------------------------------------

\subsection{Pre-assembling checks and tests}
\label{rad-test}

As both the TF-1 and F-101 lead-glass blocks have been in use for more than 14 years under 
conditions of high luminosity, there was concern about possible radiation degradation of the blocks
and the PMTs. Changes in transparency of TF-1 and F-101 lead-glasses, irradiated with 
70 GeV protons and 30 GeV $\pi^-$ mesons have been reported in \cite{Inyakin}. 
It was found that the resistance of TF-1 lead-glass against irradiation is 50 times less than that 
of F-101. An accumulated dose of 2 krad produces a degradation of transmittance of F-101 glass of
less than 1\%. It was also found that the darkening of lead-glass radiators due to irradiation
can be considerably reversed by intensive light illumination. 
Ref.~\cite{Goldberg} reports that exposure of radiation-damaged glass to UV irradiation 
or to high temperature can bring about recovery of the glass.

\begin{figure}
\begin{center}
\epsfxsize=3.40in
\epsfysize=3.40in
\epsffile{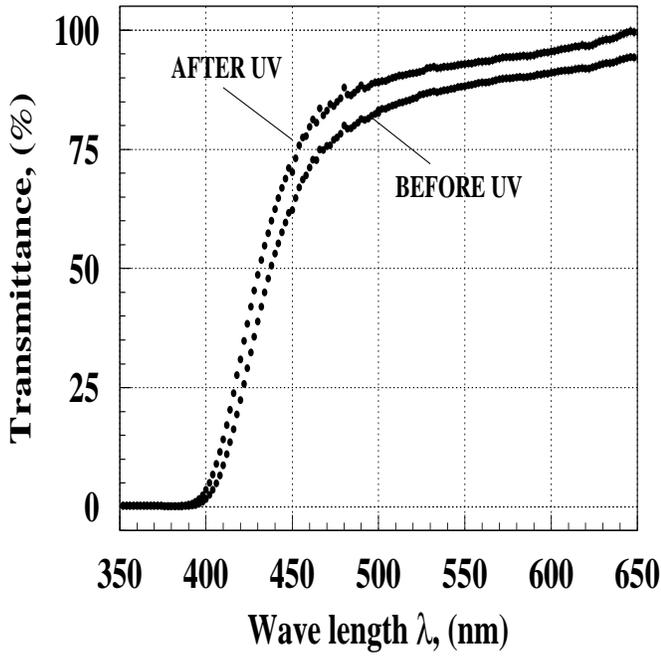}
\caption{\label{f101_rad_and_uv_trans}
Radiated ($\sim$2 krad) F-101 type lead-glass blocks transmission efficiencies 
before and after 5 days of UV curing.  
}
\end{center}
\end{figure}

The changes in transparency of TF-1 and F-101 type lead-glass radiators have been studied in
\cite{H-Mkrt10,A-Mkrt11}.
The estimated radiation dose for the used blocks was about 2 krad. 
For several samples of F-101 and TF-1 type blocks the light transmittance has been measured
before and after 5 days of curing with UV light (of wavelength $\lambda$=200-400 $nm$).
The transmission for F-101 type blocks from HERMES before and after the UV curing is
shown in Fig.~\ref{f101_rad_and_uv_trans}. We do not find significant changes in transmittance.

Note that for the TF-1 type blocks taken from the SOS calorimeter, our measurements again show
negligible degradation over more than 15 years of operation
(see Fig.~\ref{tf1_transp} in Section~\ref{tf1-block}).
This is due to efficient shielding of the SOS (HMS) spectrometer detector huts.

To summarize the results of our studies on the radiation effects, there is no evidence for
noticeable radiation damage of TF-1 and F-101 lead-glass blocks to be used in the construction
of the calorimeter for SHMS spectrometer.

\begin{figure}
\begin{center}
\epsfxsize=3.40in
\epsfysize=3.40in
\epsffile{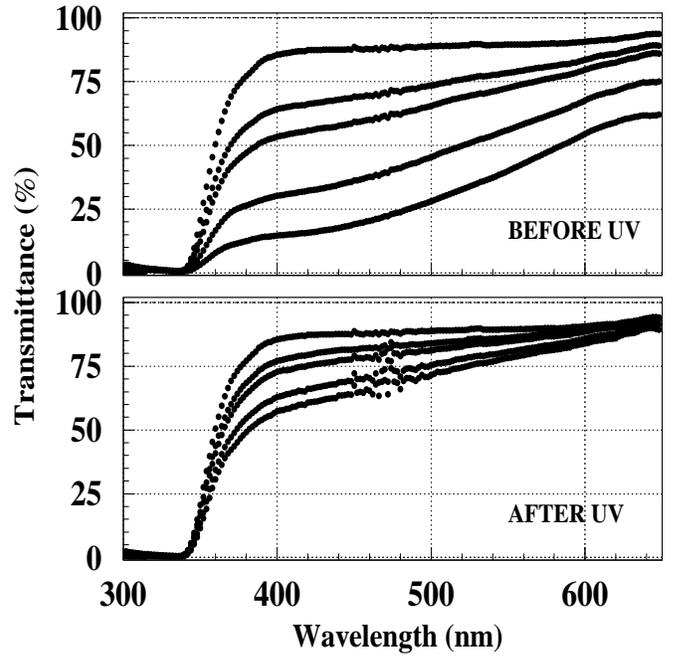}
\caption{\label{tf1_rad_and_uv_trans}
Radiated ($\sim$2-6 krad) TF-1 type lead-glass blocks' transmittance before (top panel) 
and after 5 days of UV curing (bottom panel).  
}
\end{center}
\end{figure}

As a cross check, we performed similar studies for the TF-1 type lead-glass blocks taken from
the BigCal calorimeter, which had been used in Hall C experiment Gep-III~\cite{gep3}. 
This calorimeter was operated in open geometry, and  accumulated a dose of $\sim$2-6 krad.
The results presented in Fig.~\ref{tf1_rad_and_uv_trans} show the effect of UV curing,
indicating strong radiation degradation.

The gain and relative quantum efficiencies for randomly selected PMTs from the SOS calorimeter
(XP3462B) and from the HERMES detector (XP3461) have been measured to check possible degradation 
effects in the PMTs. A simple setup with a LED light source was used to localize the Single
Electron Peak (SEP) at a given HV and define the gain for each PMT. 

\begin{figure}
\begin{center}
\epsfxsize=3.40in
\epsfysize=3.40in
\epsffile{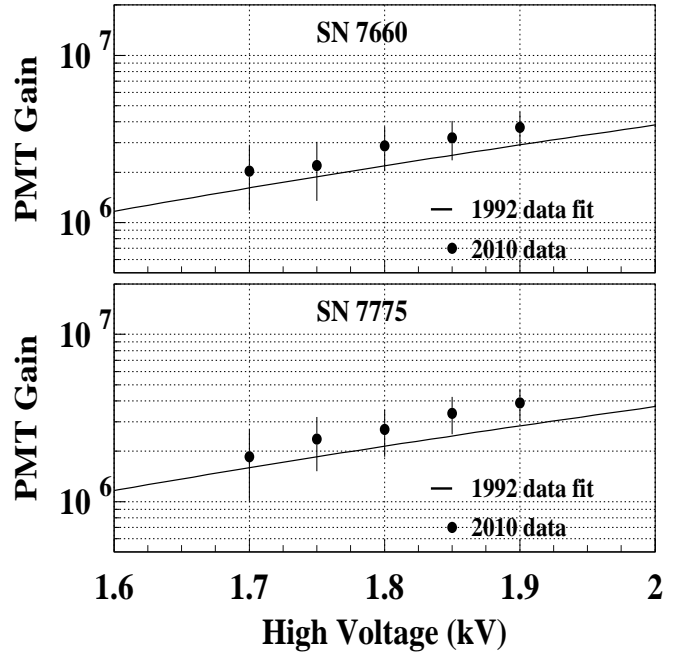} 
\caption{\label{xp3462b_gain}
Gain versus HV for two Photonis XP3462B PMTs. The lines are fits to the data from
1992 measurements. New measurements of 2010 are shown with solid symbols.  
}
\end{center}
\end{figure}

Examples of gain variation versus high voltage for the Photonis XP3462B PMT are shown in 
Fig.~\ref{xp3462b_gain}.
While 1992 and 2010 data sets agree within the errors, a systematic offset of $\sim10-15\%$
can be seen between the two, which is related to different setups used in the measurements.

For a set of PMTs dismounted from HERMES modules we have compared relative quantum efficiencies with
new XP3461 PMTs. The HV for each PMT was adjusted to the gain $\approx1.5\times10^6$.
The light intensity was adjusted to get about 100 photoelectrons from the unused new PMTs, and this
intensity was monitored by a reference PMT.
Since we kept the LED at fixed intensity and operated all the PMTs at a fixed gain, the 
difference between the detected number of photoelectrons may only come from the difference in the 
PMT QEs. The number of detected photoelectrons will in this case be directly related with the 
quantum efficiencies of PMTs.

\begin{figure}
\begin{center}
\epsfxsize=3.40in
\epsfysize=2.40in
\epsffile{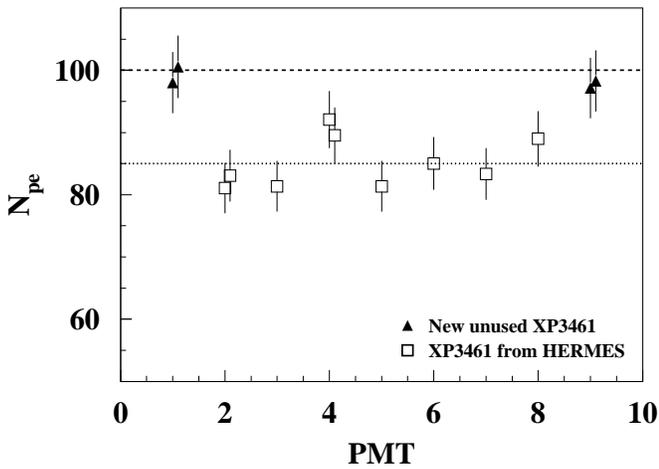}
\caption{\label{xp3461_qe_hermes_new}
Comparison of the detected number of photoelectrons at the gain 1.5$\times 10^6$ for the new XP3461
phototubes (triangle symbols), and for the PMTs taken from HERMES calorimeter (square symbols).  
}
\end{center}
\end{figure}

The comparison is shown in Fig.~\ref{xp3461_qe_hermes_new}.
A hint of aging, a 15\% systematic decrease in quantum efficiency can be noticed. However, this is 
not taken into account in the simulations, for the decrease is marginal when compared to the 
accuracy of the measurements.

%---------------------------------------------------------------------------------------------------
\subsection{Simulation code for SHMS calorimeter}
\label{geant4-code}

The code is based on GEANT4 simulation package \cite{geant4}, release 9.2. 
As in the simulations of the HMS calorimeter (see section \ref{early_mc}), the QGSP\_BERT physics
list was chosen to model hadron interactions. The code closely follows the parameters of the 
detector components mentioned in the previous sections. Other features are added into the model in 
order to bring it closer to reality as described below.

\begin{figure}
\begin{center}
\epsfxsize=3.40in
\epsfysize=2.40in
\epsffile{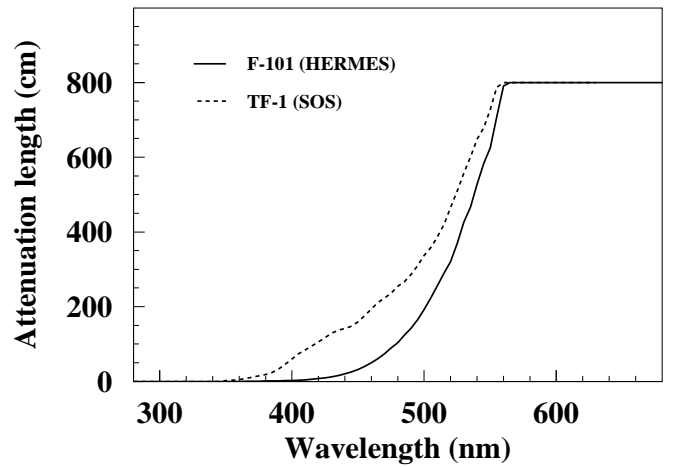} 
\caption{\label{f101_tf1_attl}
Mean attenuation lengths of F-101 (solid line) and 
TF-1 (dotted line) lead-glasses used in the simulation of SHMS calorimeter. Data below 560 $nm$ are 
extracted from transmittance measurements of the blocks. Above that the measurements are not
reliable for the extraction, and the lengths are approximated by a large constant value.  
}
\end{center}
\end{figure}

As optical measurements of both TF-1 and F-101 glasses revealed block to block variation in 
transparency, the attenuation lengths were randomly varied from block to block accordingly, around 
their mean values shown in Fig.~\ref{f101_tf1_attl}.
The quantum efficiencies of XP3462B and XP3461 PMT photocathodes are taken from the graphs provided 
by Photonis (see Fig.~\ref{xp3462b-qe} and Fig.~\ref{xp3461-qe}).
The electronic effects in data acquisition system are taken into account assuming same performance 
as for the HMS calorimeter (see section~\ref{early_mc}).

As in the HMS case, particles originate at the focal plane and traverse detector material and 
support structures in front of the calorimeter (see Table~\ref{shms-mat}).
Note, the two SHMS aerogel detectors for kaon identification \cite{E12-09-011} are not 
considered here, since their design was not finalized by the time of the calculations.
Focal plane coordinates, directions and deviations of momentum from
spectrometer setting were sampled by means of a Monte Carlo code of SHMS magnetic optics.

\begin{center}
\begin{table}
\caption{\label{shms-mat} Materials between SHMS focal plane and calorimeter that are taken into 
account in the simulation. The listed positions are at the fronts of components}
{\centering  \begin{tabular}{|c|c|c|c|c|}
\hline
Component          & Material          & position   & thickness        & density  \\
                   &                   &  (cm)      &    (cm)          & $({\rm g}/{\rm cm}^3)$\\
\hline
DC2 gas            & Ethane/Ar         &   40       &   3.81           & 0.00143   \\
DC2 foils          &  Mylar            &            & 7$\times$0.00254 & 1.4       \\
S1X hodoscope      & BC408 scint.      &   50       &    0.5           & 1.032     \\
S1Y hodoscope      & BC408 scint.      &   60       &    0.5           & 1.032     \\
Gas \v{C} gas      & $C_4F_8O$         &   80       &    109.5         & 0.0089    \\
Gas \v{C} wind.    &       Al          &            &  2$\times$0.1    & 2.6989    \\
Gas \v{C} mir.     &     glass         &            &    0.3           & 2.4       \\
Gas \v{C} mir.sup. & carbon fiber      &            &    0.1           & 1.8       \\
S2X hodoscope      & BC408 scint.      &   260      &    0.5           & 1.032     \\
S2Y hodoscope      &    Quartz         &   265      &    2.5           & 2.634     \\
Preshower sup.     &     Al            &   269      &    0.05          & 2.6989    \\
Preshower cov.     &     Al            &   280      &    0.1           & 2.6989    \\
Shower sup.        &     Al            &   282      &    0.05          & 2.6989    \\
\hline
\end{tabular}\par}
\end{table}
\end{center}

Light tracing is done within the frame of GEANT4 optics model. All the components related to the 
tracking of optical photons --- like lead glass blocks, reflective foil wrapper, air layer between
the reflector and the block, PMT glass windows, optical couplings of the windows and the blocks --- 
were coded in terms of their sizes and optical parameters.

The calibration algorithm  used in these studies is the same as for the HMS calorimeter 
(see subsection~\ref{calo_calibr}): the variation of total energy deposition in Preshower and Shower
relative to the energy of the primary electron is minimized with respect to the calibration 
constants for each signal channel.
The signals from Preshower are corrected for the horizontal coordinate of impact point. The signals 
from Shower are not corrected for impact point coordinates.

%---------------------------------------------------------------------------------------------------
\subsection{Performance of SHMS calorimeter}
\label{shms-calo-perf}

\begin{figure}
\begin{center}
\epsfxsize=3.40in
\epsfysize=2.40in
\epsffile{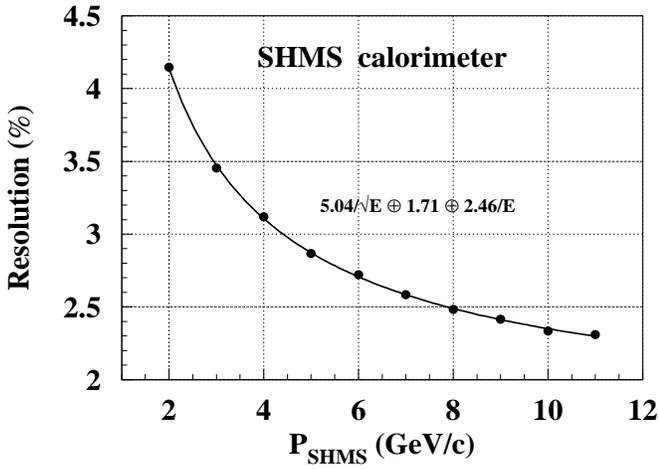}
\caption{\label{shms_calo_res} Resolution of the modeled SHMS calorimeter. The small bullet symbols 
are data from the GEANT4 simulation, the line is the conventional 3-parameter fit \cite{pdg}
to them.  
}
\end{center}
\end{figure}

Resolution of the modeled SHMS calorimeter (Fig.~\ref{shms_calo_res}) is analogous to what
has been reported for other lead-glass shower counters (references 21 through 31 in 
\cite{Tadevos2010}), though it is somewhat lower compared to the HMS calorimeter
(compare Fig.~\ref{shms_calo_res} with Fig.~\ref{hcal_sigma}).
Examination of the functional forms of energy dependencies of the two resolutions shows that the 
difference comes mainly from the stochastic term: compare 5.04$\%\sqrt{E}$ for the SHMS with 
3.75$\%\sqrt{E}$ for the HMS. 
The stochastic term is sensitive to dead material before detector and to photoelectron
statistics~\cite{pdg}, which is in turn sensitive to the quality of radiator and light detectors.
Both of these conditions are less favorable for the SHMS counter: there is more material between
the focal plane and the calorimeter in the SHMS than in the HMS $-\sim$0.38 versus
$\sim$0.16 radiation lengths respectively; and the lead-glass in SHMS calorimeter is less 
transparent than in the HMS calorimeter. The latter, combined with larger sizes, noticeably reduces 
photoelectron statistics in the SHMS calorimeter.

\begin{figure}
\begin{center}
\epsfxsize=3.40in
\epsfysize=3.40in
\epsffile{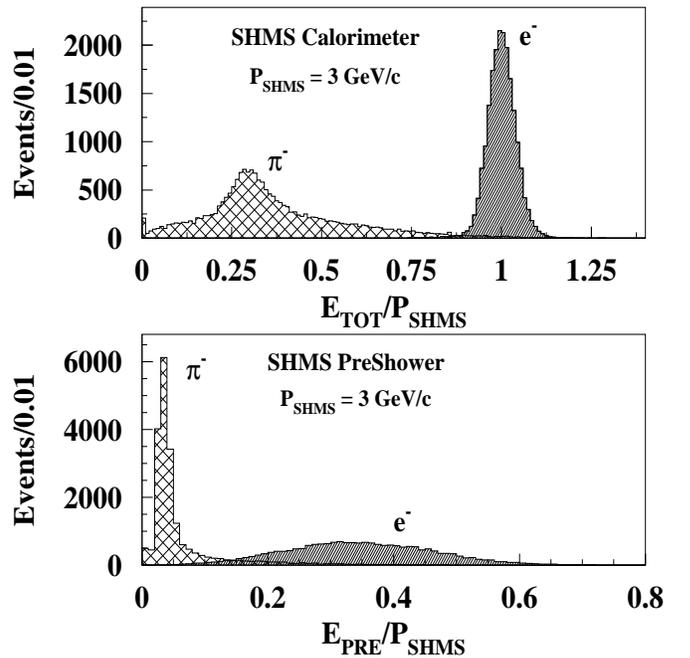}
\caption{\label{shms_calo_endep}
Distribution of the normalized energy deposition from pions (hatched area) and 
electrons (full histogram) in SHMS calorimeter as a whole (top) and in Preshower only (bottom) at 
3 GeV/c momentum setting.  
}
\end{center}
\end{figure}

Despite that, decent electron/hadron separation can be achieved
by using the signal from the Preshower in addition to the total energy deposition in the calorimeter.
As an illustration of hadron/electron rejection capability, example histograms of energy depositions
from $e^-$ and $\pi^-$ in the calorimeter and in the Preshower are presented in
Fig.~\ref{shms_calo_endep}.
As it is seen in the bottom panel, the minimum ionizing pions and the 
showering electrons are separable to some extent in Preshower.

\begin{figure}
\begin{center}
\epsfxsize=3.40in
\epsfysize=3.40in
\epsffile{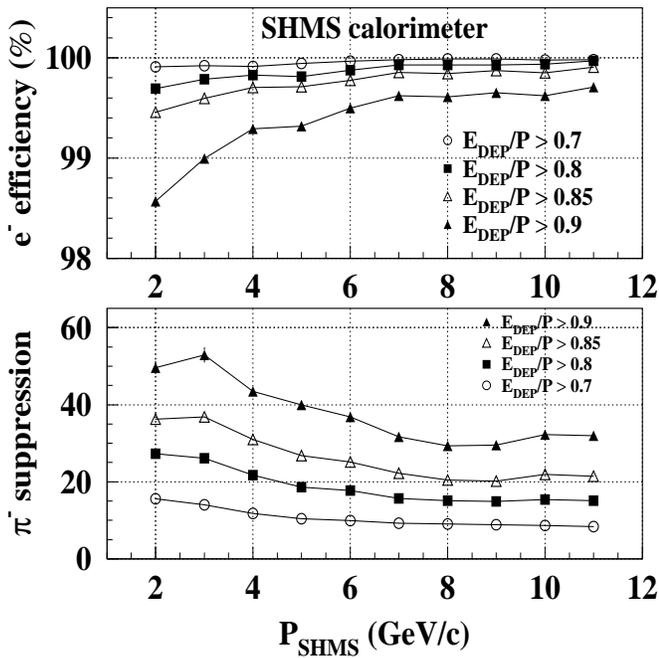}
\caption{\label{shms_calo_eff}
Electron detection efficiency (top) and pion suppression factor (bottom) of the SHMS calorimeter
versus spectrometer's momentum setting for different cuts on the normalized total energy deposition.
}
\end{center}
\end{figure}

Electron detection efficiency and pion suppression factor for different cuts on the normalized 
total deposited energy are shown in Fig.~\ref{shms_calo_eff} (compare with Fig.~\ref{hms_eff_vs_p}
and Fig.~\ref{hms_cut_sup_vs_p} for HMS calorimeter).
For a constant cut, $e^-$ detection improves with momentum, which is consistent
with better resolutions at higher energies. Meanwhile, $\pi^-$ rejection tends to worsen
because of the increase in electromagnetic component of hadron induced cascades.
The cut $E_{Dep}/P>0.7$ ensures $e^-$ detection better than 99.8\% but modest $\pi^-$ 
suppression of $\sim$10. By imposing higher cuts one can trade off $e^-$ detection efficiency for a 
higher $\pi^-$ suppression.

When compared to the HMS calorimeter, at the same cuts on total deposited energy the SHMS
calorimeter ensures somewhat better $e^-$ detection efficiency due to lower fraction of events
of low visible energy deposition.
Meanwhile, the $\pi^-$ suppression is noticeably decreased (compare bottom panel in
Fig.~\ref{shms_calo_eff} with Fig.~\ref{hms_cut_sup_vs_p}).

Calorimeter segmentation allows one to take advantage of the differences in the space
development of electromagnetic and hadronic showers for PID. Electromagnetic showers develop
earlier and deposit more energy at the start than hadronic cascades.
Thus measuring energy deposited in the front layer of a detector along with
total energy deposition improves the electron/hadron separation.

\begin{figure}
\begin{center}
\epsfxsize=3.40in
\epsfysize=3.40in
\epsffile{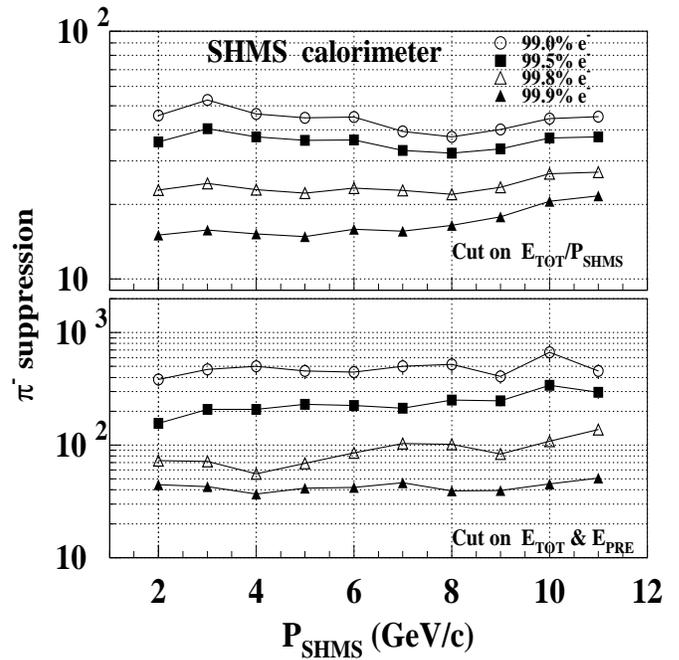}
\caption{\label{shms_calo_sup_p}
Pion suppression factor versus SHMS momentum setting obtained with the two PID methods,
for the different electron detection efficiencies indicated on the top panel. 
Data on the top panel are obtained by cutting on the normalized total energy deposition. 
Data on the bottom panel are obtained by applying two-dimensional cuts on the combination of total 
energy deposition and in the Preshower only.  
}
\end{center}
\end{figure}

\begin{figure}
\begin{center}
\epsfxsize=3.40in
\epsfysize=3.40in
\epsffile{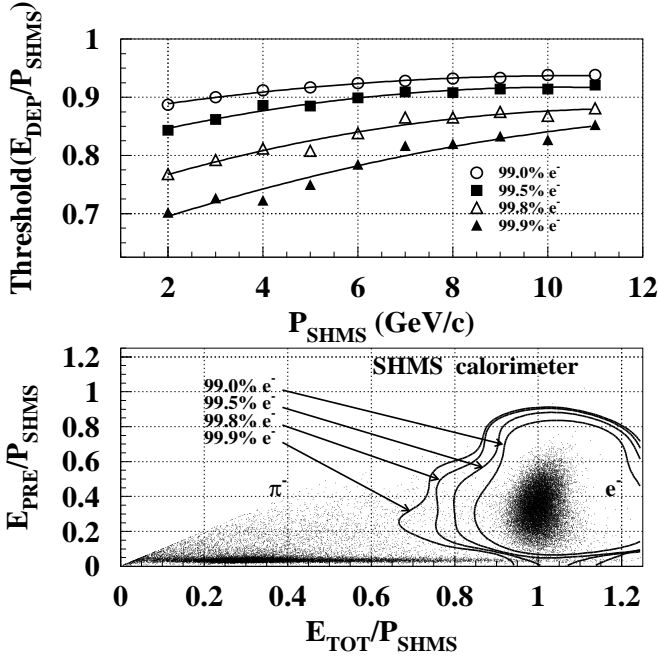}
\caption{\label{shms_calo_cuts}
Top: Cuts on the normalized total energy deposition for different $e^-$ detection efficiencies. 
The quadratic fits to the data points have been used for evaluation  of $\pi^-$ suppression factors 
shown on the top panel of Fig.~\ref{shms_calo_sup_p}.   
Bottom: An example $e^-/\pi^-$ separation by using combination of the normalized energy deposition
in the calorimeter (Preshower+Shower) and Preshower at 3 GeV/c SHMS momentum setting. 
(In on-line electron events are in red, pion events are in blue). 
The loops are boundary for different $e^-$ detection efficiencies, optimized by means of SVM 
neural network.
}
\end{center}
\end{figure}

Pion suppression with the two PID methods $-$ by using total energy deposition alone, and energy
deposition in the Preshower together with total energy deposition $-$ are compared in
Fig.~\ref{shms_calo_sup_p}.
Suppression factors on the top panel are obtained by imposing cuts on the total deposited
energy. The cuts (shown in the top panel of Fig.~\ref{shms_calo_cuts}) are chosen to ensure 
the electron detection efficiencies listed in the figures. 

The suppression factors on the bottom panel are obtained by separation of pion and electron events
of concurrent energy depositions in the Preshower and in the whole calorimeter (exemplified in the 
Fig.~\ref{shms_calo_cuts}, bottom panel). 
The separation boundaries are tuned to the same electron detection efficiencies as in the first case,
and are optimized for minimum error rate by means of SVM$^{light}$ neural network~\cite{svm_light}.
Details can be found in a similar case with HMS calorimeter ~\cite{Tadevos2010}, where the forward
layer of the counter was used as preshower. There, for the PID with combined energy depositions,
from comparison with experimental data it was found that the simulation overestimates pion
suppression, by $\sim$70\% at low electron detection efficiencies $\gtrsim$90\%, and
$\sim$40\% at high efficiencies $\sim$99.7\%.

\begin{figure}
\begin{center}
\epsfxsize=3.40in
\epsfysize=2.40in
\epsffile{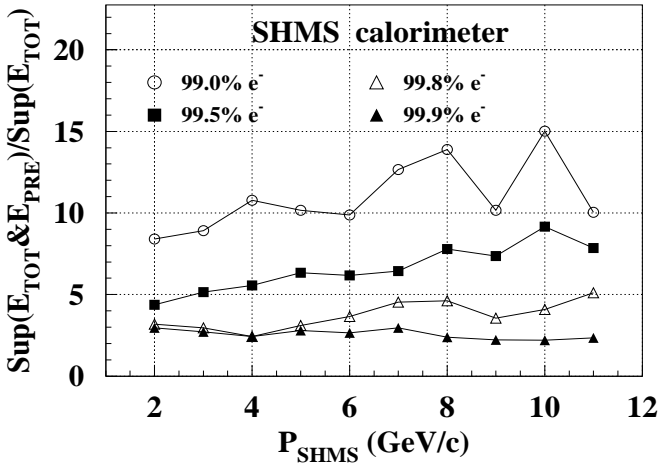}
\caption{\label{shms_calo_sup_gain} 
Gain in $\pi^-$ suppression from using energy deposition in 
Preshower along with total deposited energy in SHMS calorimeter.  
}
\end{center}
\end{figure}

As it is seen in Fig.~\ref{shms_calo_sup_p}, in both cases there is a trend that suggests improvement
of the $\pi^-$ rejection with increase of momentum.
Combining the total energy deposition $E_{tot}$ with deposition in the Preshower $E_{pre}$
significantly improves pion rejection. Gain in suppression by a factor of 2 - 10 times is
achievable, dependent on momentum and the chosen e- efficiency.
Generally the gain is bigger at higher momenta and for lower $e^-$ detection efficiencies 
(see Fig.~\ref{shms_calo_sup_gain}).
Even for very high electron efficiencies, the combined cut yields a factor of two or more
improvement in the pion rejection over the simple E$_{TOT}$ cut.
By using the Preshower the PID capabilities of the SHMS calorimeter become
as good as that of HMS calorimeter where the first layer serves as Preshower. 

To summarize results on the SHMS calorimeter, the GEANT4 simulations were conducted with realistic 
parameters of the detector. The simulations predict a resolution similar to other lead-glass 
counters, though somewhat worse than for the existing HMS calorimeter.
Good electron/hadron separation 
can be achieved by using energy deposition in the Preshower along with total energy deposition in 
the calorimeter. In this case the PID capability is similar to the one attainable with the HMS 
calorimeter. A pion suppression factor of a few hundred is predicted at 99\% electron efficiency.

%%%%%%%%%%%%%%%%%%%%%%%%%%%%%%%%%%%%%%%%%%%%%%%%%%%%%%%%%%%%%%%%%%%%%%%%%%%%%%%%%%%%%%%%%%%%%%%%%%%%

\section{SUMMARY AND CONCLUSIONS}

In summary, we have developed and constructed electromagnetic calorimeters from TF-1 type lead-glass
blocks for the HMS and SOS magnetic spectrometers at JLab Hall C. The energy resolution better than
$\sigma/E \sim 6\% /\sqrt E$ and the pion suppression $\sim$100:1 for $\sim$99\% $e^-$
detection efficiency have been achieved in the 1 -- 5 GeV energy range.
Performance of the HMS calorimeter within full momentum range of the spectrometer, attainable after
CEBAF 12 GeV upgrade, is modeled by GEANT4 simulation.
Within the limited momentum range the calculated resolution and $\pi^-$ suppression factor are
in good agreement with experimental data.
The simulated pion suppression systematically exceeds experiment, by less than a factor of two,
which is acceptable for rejection studies. The HMS/SOS calorimeters have been used in nearly all
the Hall C experiments, providing good energy resolution and high pion suppression factor.
No significant deterioration in the performance is observed in the course of operation
since 1994.

Design construction of the electromagnetic calorimeter for the newly built SHMS spectrometer in
Hall C has been finalized, based on extensive exploratory studies.
From a few considered versions, the Preshower+Shower configuration  was selected as most
cost-effective.
The Preshower will consist of a layer of 28 modules with TF-1 type lead glass radiators,
stacked back to back in two columns.
The Shower part will consist of 224 modules with F-101 type lead glass radiators, stacked in a
``fly eye'' configuration of 14 columns and 16 rows.
$120\times130~{\rm cm}^2$ of active area will cover beam envelope at the calorimeter.

A Monte Carlo program for the newly designed SHMS shower counter was developed, based on the GEANT4
simulation package, and simulations have been conducted with realistic parameters of the detector. 
The predicted resolution yields somewhat to the HMS calorimeter.
Good electron/hadron separation can be achieved by using energy deposition in the Preshower along 
with total energy deposition in the calorimeter. In this case the PID capability is similar to
or better than those attainable with HMS calorimeter.
A pion suppression factor of a few hundreds is predicted for 99\% electron detection efficiency.

%%%%%%%%%%%%%%%%%%%%%%%%%%%%%%%%%%%%%%%%%%%%%%%%%%%%%%%%%%%%%%%%%%%%%%%%%%%%%%%%%%%%%%%%%%%%%%%%%%%%
\vspace{0.4in}
\centerline{ACKNOWLEDGMENTS}

The authors wish to thank Tsolak~Amatuni for the work on hardware and software of
HMS/SOS calorimeters in the development and construction stages, and for the idea
of using support vector machines for particle identification with segmented calorimeters.
We thank Ashot~Gasparyan for the work and fruitful ideas in the early stages of the hardware
development.
We thank Carl~Zorn from Detector Group of the Physics Division for outstanding support during
optical studies of the lead-glass blocks and PMTs.

We thank William~Vulcan, Joseph~Beaufait and Hall C technical staff for helping in all areas of
preparation, assembling and installation of the detectors in HMS and SOS huts, mounting electronics
and  cable communications for HMS/SOS calorimeters.

This work is supported in part by ANL grant DE-AC02-06CH11327.

The Southeastern Universities Research Association operates the Thomas Jefferson National 
Accelerator Facility under the U.S. Department of Energy contract DEAC05-84ER40150.

%%%%%%%%%%%%%%%%%%%%%%%%%%%%%%%%%%%%%%%%%%%%%%%%%%%%%%%%%%%%%%%%%%%%%%%%%%%%%%%%%%%%%%%%%%%%%%%%%%%%

%%%%%%%%%%%%%%%%%%%%%%%%%%%%%%%%%%%%%%%%%%%%%%%%%%%%%%%%%%%%%%%%%%%%%%%%%%%%%%%%%%%%%%%%%%%%%%%%%%%

\end{document}